\documentclass[useAMS,usenatbib]{mn2e}
\usepackage{amssymb}
\usepackage{natbib}
\usepackage{times}
\usepackage{graphics,epsfig}
\usepackage{graphicx}
\usepackage{amsmath}
\usepackage{amssymb}
\usepackage{comment}
\usepackage{supertabular}

\usepackage{graphicx}
\usepackage{amsmath}
\usepackage{color}

\newcommand{\kms}{km~s$^{-1}$}
\newcommand{\arcs}{$^{\prime\prime}$}

\newcommand{\shi}{\ensuremath{\Sigma_{\rm HI}}}

\newcommand{\ms}{\ensuremath{\rm M_{\odot}}}

\newcommand{\acc}{\ensuremath{\rm atoms~cm^{-2}}}

\newcommand{\mspcc}{\ensuremath{\rm M_{\odot} pc^{-3} }}
\newcommand{\mspc}{\ensuremath{\rm M_{\odot} pc^{-2} }}
\newcommand{\HI}{H{\sc i}}

\begin{document}

\title {A self-consistent hydrostatic mass modelling of pressure supported dwarf galaxy Leo T}

\author [N. N. Patra]{	Narendra Nath Patra$^{1}$ \thanks {E-mail: narendra@ncra.tifr.res.in} \\
	$^{1}$ National Centre for Radio Astrophysics, Tata Institute of Fundamental Research, Pune University campus, Pune 411 007, India\\	
}

\date {}
\maketitle

\begin{abstract}

Assuming a hydrostatic equilibrium in an \HI~cloud, the joint Poisson's equation is set up and numerically solved to calculate the expected \HI~distribution. Unlike previous studies, the cloud is considered to be non-isothermal, and an {\it iterative} method is employed to iteratively estimate the intrinsic velocity dispersion profile using the observed second-moment of the \HI~data. We apply our {\it iterative} method to a recently discovered dwarf galaxy Leo T and find that its observed \HI~distribution does not comply with the expected one if one assumes no dark matter in it. To model the mass distribution in Leo T, we solve the Poisson's equation using a large number of trial dark matter halos and compare the model \HI~surface density ($\Sigma_{HI}$) profiles to the observed one to identify the best dark matter halo parameters. For Leo T, we find a pseudo-isothermal halo with core density, $\rho_0 \sim 0.67$ \mspcc~and core radius, $r_s \sim 37$ parsec explains the observation best. The resulting dark matter halo mass within the central 300 pc, $M_{300}$, found to be $\sim 2.7 \times 10^6$ \ms. We also find that a set of dark matter halos with similar $M_{300} \sim 3.7 \times 10^6$ \ms~ but very different $\rho_0$ and $r_s$ values, can produce equally good $\Sigma_{HI}$ profile within the observational uncertainties. This, in turn, indicates a strong degeneracy between the halo parameters and the best fit values are not unique. Interestingly, it also implies that the mass of a dark matter halo, rather than its structure primarily directs the expected \HI~distribution under hydrostatic equilibrium.

\end{abstract}

\begin{keywords}
ISM: clouds -- ISM: structure -- galaxies: dwarf -- galaxies: ISM -- galaxies: individual: Leo T -- (cosmology:) dark matter
\end{keywords}

\section{introduction}

Though the $\rm \Lambda-CDM$ model of cosmology has been immensely successful in explaining the observable universe at large scales \citep{efstathiou92,riess98,spergel07}, several inconsistencies yet persist between its predictions and observations in smaller scales \citep{boylankolchin11,boylankolchin12,hayashi12}. The `missing satellite' problem \citep{klypin99,moore99,wadepuhl11} poses one of the significant challenges to the current understanding of the $\rm \Lambda-CDM$ model. The number of predicted small galaxies within the virial radii of massive galaxies, (for example the Milky Way) are found to be an order of magnitudes higher than what is observed as luminous satellites of such galaxies.

Several recent numerical and observational efforts have closed the gap between the predicted and observed Milky Way satellites significantly though \citep[][]{sand15,zolotov12,kravtsov04,faerman13,belokurov07}, the difference remains. Amongst several solutions proposed for this problem, HVC-minihalo hypothesis is one of the promising one \citep{sternberg02,giovanelli10,faerman13}. In this hypothesis, the population of unidentified satellite mini-halos are camouflaged as the Compact High-Velocity Clouds or Ultra Compact High-Velocity Clouds (CHVCs/UCHVCs) observed in the circumgalactic medium of the Galaxy. As these satellite galaxies do not have enough observable star formation, very often they are referred as dark galaxies. 

The star formation recipes in these low mass satellites are not well understood and deviates \citep{roychowdhury09,roychowdhury14} from the well-established prescriptions for larger galaxies \citep{kennicutt98b,leroy13}. This, in turn, leads to a poor assessment of the luminosity function and subsequently the number counts of faint satellites which are expected to be detected in an optical survey. In fact a number of attempts to detect stellar component in potential dark galaxy candidates resulted in null detections \citep{hopp03,hopp07,siegel05,willman02,davies02,simon02,sand15}. However, the poor understanding of star formation laws in these smallest galaxies are not expected to affect the \HI~mass function; and these low mass objects are expected to be detected in adequate numbers in large area \HI~surveys. There already exist a number of observational efforts detecting these potential dark galaxies \citep{adams16,adams15,oosterloo13,giovanelli10,irwin07,rhode13}, and many more are expected to be detected in future large area \HI~surveys (e.g. Medium deep Survey by APERTIF, WALLABY by ASKAP \citep{duffy12} etc.).

However, the \HI~properties of these galaxies are expected to be very similar to that of the CHVCs or UCHVCs. This, in turn, would make it difficult to identify the possible satellite galaxies from a pool of CHVCs/UCHVCs just by looking at their \HI~morphology \citep[see, e.g.][]{sternberg02}. However, a galaxy with cosmological origin will be hosted by a dark matter halo whereas an \HI~cloud of galactic origin would not. As a self-gravitating cloud can not induce rotation by itself, the presence of a rotating disk in a galaxy is considered to be a reliable indicator of the presence of a dark matter halo in it. However, though, very often the smallest galaxies lack enough angular momentum to induce rotation in their discs \citep[see, e.g.,][]{adams16,adams15}. In these situations, measurement of the rotation curve is not possible. 
 
Along with detecting potential satellite galaxies, measurement of the dark matter distribution in these galaxies would provide critical constraints to the cosmological models of structure formation. For example, \citet{strigari08} found a common mass scale in MW satellite galaxies by performing a maximum likelihood analysis on the line-of-sight velocity dispersion data of stars. They found that all the satellite galaxies have a similar amount of dark matter mass of $\sim 10^7$ \ms within their central 300 pc. Hence, it would be interesting to perform mass modelling of these smallest galaxies to investigate their dark matter distribution. However, the inability to measure the rotation curve in these galaxies inhibits the mass-modelling in a conventional way.

Both dark galaxies and clouds that originate from the galactic ISM are likely to be in rough hydrostatic equilibrium; however, only dark galaxies are expected to have a dark matter halo. The presence of the dark matter halo is expected to affect the distribution and kinematics of the \HI~gas. Conversely, the observed \HI~distribution, along with the assumption of hydrostatic equilibrium could be used to constrain the amount of dark matter associated with an \HI~cloud. Here we build a detailed hydrostatic equilibrium model for Leo T, a recently discovered dwarf galaxy. However, this modelling can be used for any pressure supported galaxy to estimate its dark matter content and distribution.

Leo T is one of the smallest known gas-rich galaxies in the local universe and is at an estimated distance of only $\sim$420 kpc \citep{irwin07}. The \HI~morphology of this galaxy is very similar to that of the CHVCs/UCHVCs \citep{ryan08}. This makes it an excellent test bed for using hydrostatic equilibrium models to try and distinguish between dwarf galaxies and UCHVCs/CHVCs. Previous studies, e.g., \citet{faerman13} developed models for gas-rich objects, where hydrostatic equilibrium along with photoionisation sets the \HI~half-mass radius. These models are not self-consistent as they do not account for the gravity of baryons. Not only that, but these models also assume the galaxy to be a Warm Neutral Medium (WNM) cloud with a fixed temperature of $\sim 10^4$ Kelvin. However, the interstellar medium (ISM) of galaxies is known to be dominated by turbulent motions which set the equilibrium in the neutral phase of the ISM \citep{young96,younglo97,young03}. In this scenario, a thermal temperature of $10^4$ Kelvin might not be a good proxy to the hydrostatic pressure.

In this paper, we set up self-consistent (i.e., accounting for the gravity of the gas and stars) models for \HI~clouds with or without dark matter. We also try a variety of different dark matter halos and determine via a Monte-Carlo approach the best dark matter halo which produces the observed \HI~distribution for Leo T. This allows us to perform a mass-modelling of Leo T providing constraints on parameters like central dark matter density as well as the core radius of the isothermal dark matter halos that we use for the modelling.

\section{Formulation of equation and constraints}

Here we consider \HI~clouds which are very similar in \HI~properties to that of small pressure supported galaxies (e.g., Leo T) or the UCHVCs/CHVCs. We also consider that these clouds reside in the circumgalactic medium with a nominal distance of $\sim$ 1 Mpc. In the absence of any large-scale motion or rotation, these \HI~clouds then can be assumed to be in approximate hydrostatic equilibrium. In fact, the circular symmetry observed in resolved \HI~maps of UCHVCs/CHVCs \citep{adams13} or in Leo T \citep{adams18}, the only known pressure supported galaxy indicates a minimal presence of large-scale motion or external disturbances. The lack of rotation in these objects can also be confirmed from the observed \HI~velocity field. Lastly, these objects are highly unlikely to be confined by the external pressure of Hot Ionised Medium (HIM) or radiation. As shown in details in \citet{faerman13}, a pressure of $P_{th}/k_b = n_H T \gtrsim 200 d^{-1}_{Mpc} cm^{-3} K$ is required to support a typical UCHVC at a nominal distance of $\sim$ 1 Mpc. Where $n_H$ is the volume density of \HI~in a typical CHVC/UCHVC and $d$ is the distance in Mpc. $T$ denotes the temperature of the HIM (~few times $10^6$ K). This, in turn, demands such a dense HIM that its total mass inside 1 Mpc would be comparable to the dynamical mass of the entire local group. Hence, pressure confinement of the UCHVCs/CHVCs seems to be highly unlikely. Given these, it is reasonable to assume these \HI~clouds to be in hydrostatic equilibrium.

The Poisson's equation for a spherically symmetric system then can be written as,
\begin{equation}
\nabla^2 \Phi = 4 \pi G \left( \rho_{s} + \rho_{g} + \rho_{dm} \right)
\label{eq1}
\end{equation}

where $\Phi$ is the total gravitational potential and $\rho_s$, $\rho_g$ and $\rho_{dm}$ are the volume density of stars, gas and dark matter halo respectively.

As we are interested in neutral \HI~clouds, we expect the amount of radiation and pressure due to it would be negligible and can be ignored in the formulation of hydrostatic equilibrium equation. In that case, the pressure will be entirely due to thermal or random turbulent motions which will be balanced by the gradient in the gravitational potential.

\begin{equation}
\frac{1}{\rho} \frac{dP}{dR} + \frac{\partial \Phi}{\partial R} = 0 
\label{eq2}
\end{equation}

Assuming an ideal gas equation, the pressure could be written in terms of velocity dispersion which is an observable (indirect) quantity.
\begin{equation}
P = \rho \textless \sigma^2 \textgreater
\label{eq3}
\end{equation}

Combining equations (\ref{eq1}), (\ref{eq2}) and (\ref{eq3}), one can eliminate $\Phi$ and $P$ to obtain

\begin{equation}
\begin{aligned}
\frac {d^2 \rho} {dR^2} = {} & ~~~~\left[\frac{1}{\rho}\left(\frac{d\rho}{dR}\right)^2 - \left( \frac{2}{R} + \frac{2}{\sigma} \frac{d\sigma}{dR}\right)\frac{d\rho}{dR}\right] \\
& - \left[\frac{2\rho}{\sigma^2} \left(\frac{d\sigma}{dR}\right)^2 + \frac{4\rho}{\sigma R} \frac{d\sigma}{dR} + \frac{2\rho}{\sigma} \frac{d^2 \sigma}{dR^2} \right] \\
&- \left[\frac{4 \pi G \rho}{\sigma ^2}\left(\rho_{s} + \rho_{g} + \rho_{dm}\right)\right]
\end{aligned}
\label{eq4}
\end{equation}

\noindent Here, $\sigma$ represents the actual velocity dispersion at any radius {\it r}. The above equation cannot be solved without the knowledge of velocity dispersion ($\sigma$) as a function of radius. But, $\sigma$ is not a directly measurable quantity. In \HI~spectral line observations, the intensity weighted velocity dispersion, i.e., the second moment (M2) is measured along a line of sight instead of $\sigma$. To overcome this, we construct a procedure called {\it iterative} method in which we estimate the correct $\sigma$ profile by using the observed M2 profiles to solve Eq.~\ref{eq4}. We describe this method in details in \S~\ref{itr_method}.

\begin{figure}
\begin{center}
\begin{tabular}{c}
\resizebox{0.45\textwidth}{!}{\includegraphics{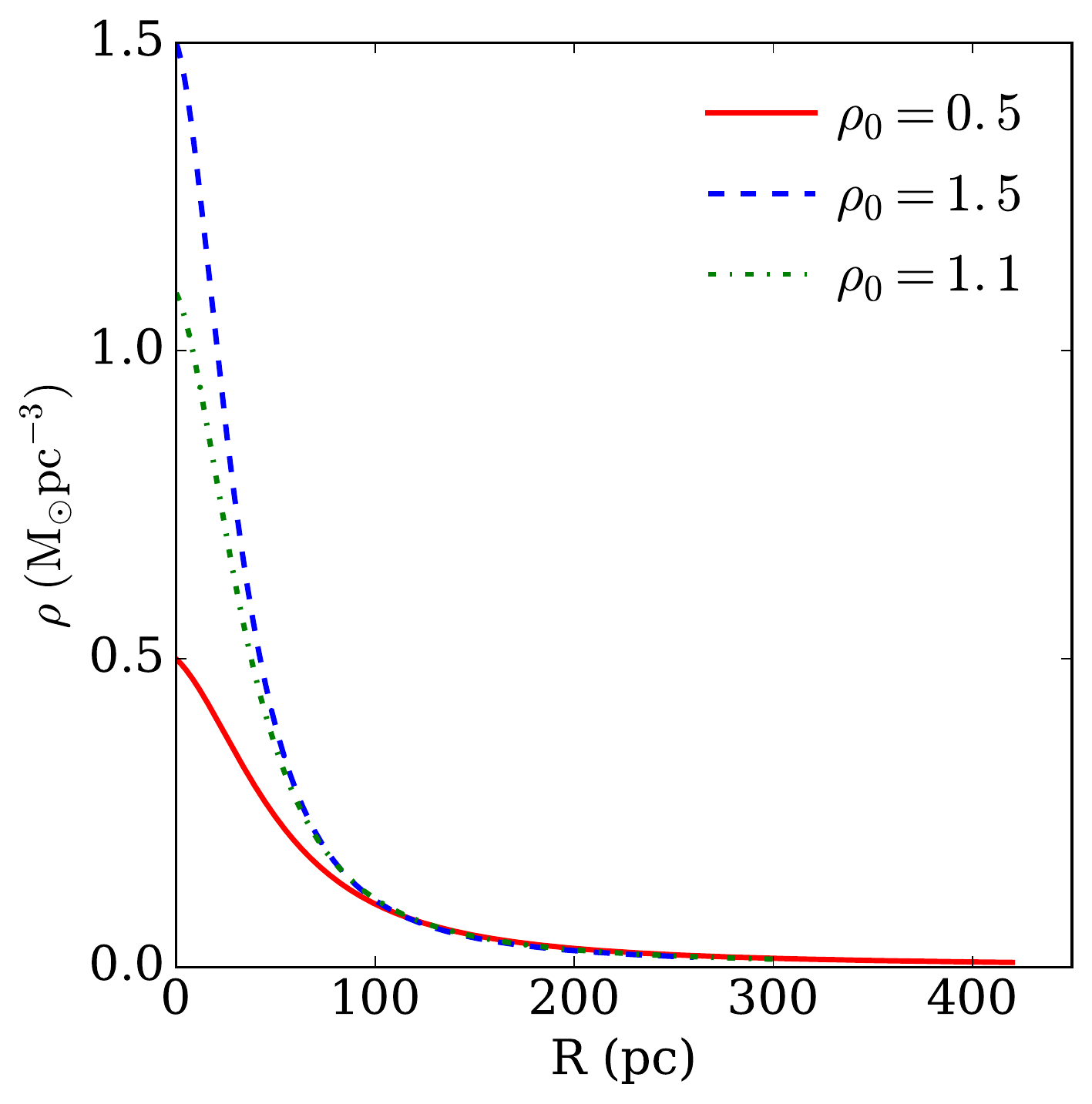}} \\
\resizebox{0.45\textwidth}{!}{\includegraphics{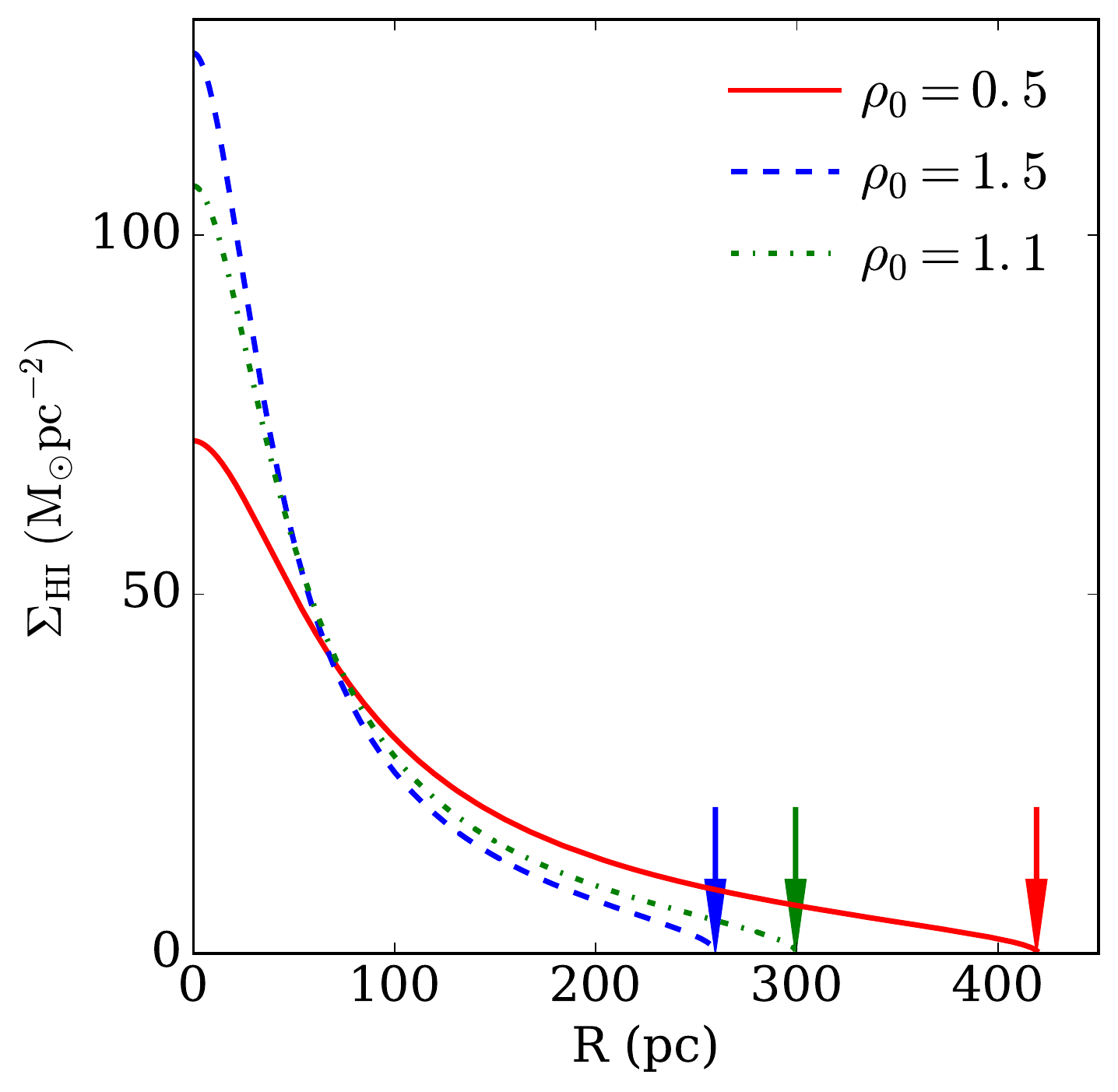}} 
\end{tabular}
\end{center}
\caption{The example solutions of Eqn.~\ref{eq4} for different central \HI~densities, $\rho_0$. Top panel: The density solutions for different $\rho_0$ values. Bottom panel: Corresponding \shi~profiles. The arrows indicate the observed \HI~size of the clouds for a 3-sigma column density sensitivity of $0.12~M_{\odot} pc^{-2}$ ($1.5 \times 10^{19}$ \acc).}
\label{exaple_soln}
\end{figure}

Eq.~\ref{eq4} is a second-order partial differential equation which we solve numerically using 8$^{th}$ order Runge-Kutta method as implemented in python package {\tt scipy}. As it is a second-order differential equation, we need at least two initial conditions to solve it. In hydrostatic equilibrium, the density at the centre would be maximum leading to $\frac{d\rho}{dR}=0$ at $R=0$. However, $\rho$ at $R=0$, i.e., $\rho_0$ is not a directly measurable quantity, instead, it can be guessed such as to produce correct observed \HI~distribution. In Figure~\ref{exaple_soln} top panel, we plot example solutions for different assumed $\rho_0$. The bottom panel of the figure shows resulting \HI~surface densities. To obtain these \HI~surface densities a 3D model of the cloud is built using the example solutions and convolved with an assumed telescope beam of 15\arcs $\times$ 15\arcs~after projecting it to the two-dimensional sky plane. An observed 3-sigma column density threshold is applied to determine the \HI~cloud size or maximum \HI~radius. As can be seen from the bottom panel of Fig.~\ref{exaple_soln}, different central \HI~densities produce different maximum \HI~radius as marked by the arrows.

To solve Eq.~\ref{eq4} for a particular \HI~cloud, estimating the right central \HI~density which produces the observed \HI~cloud size is essential. We do it in an iterative manner adopting a bi-section approach. We pick two central \HI~densities in a trial and error basis which produces \HI~cloud size such that it encloses the observed \HI~cloud size. For example, in Fig.~\ref{exaple_soln} bottom panel, we used two central densities as 0.5, and 1.5 \mspcc~which produce \HI~clouds of sizes $\sim$ 420 pc and $\sim$ 260 pc (for a 3-sigma column density of 0.12 \mspc) respectively (marked with red and blue arrows). If we assume a cloud size of say 300 pc, adopting a bisection method, the final central density can be narrowed down such that it produces the observed cloud size with an accuracy better than one percent. In this example it is found to be 1.1 \mspcc. The bisection method is found to converge quickly within 10s of iterations if the initial central densities are assumed reasonably. 
 
The intrinsic velocity dispersion profile is one of the important input parameters in equation~\ref{eq4} and should be known a priory to solve it. In Fig.~\ref{exaple_soln}, we have used a linear intrinsic velocity dispersion profile of the form $\sigma (R) = 0.012\times R + 4$ for demonstration, where $\sigma$ is in km/s and $R$ is in parsec. However, practically, $\sigma$ is not a directly measurable quantity and in the following section, we develop an {\it iterative} method to estimate the $\sigma$ profile using the observed M2 profile and solve Eq.~\ref{eq4}.

\section{Iterative method}
\label{itr_method}

To solve Eq.~\ref{eq4} and calculate the \HI~distribution in a cloud, one primarily needs two inputs, the central \HI~density and the $\sigma$ profile. But none of these quantities is directly observable. The central \HI~density is indirectly determined by the observed cloud size. Whereas the $\sigma$ profile is estimated using the {\it iterative} method. In this method, we try to estimate a correct intrinsic $\sigma$ profile which in hydrostatic equilibrium, will simultaneously produce the observed \HI~cloud size and the observed M2 profile. We start with the observed M2 ($M2_{obs}$) profile as the input $\sigma$ profile ($\sigma_{in}$) and successively decrease it in minute steps (in each iteration) such that we achieve a $\sigma_{in}$ which is now capable of producing the observables within the allowed uncertainties.

To do that, in the first iteration, we solve Eq.~\ref{eq4} using the $M2_{obs}$ as $\sigma_{in}$. Consecutively, using the solutions, i.e., $\rho(R)$, we build a 3D model of the cloud. Using this 3D model, we produce an \HI~spectral cube and the corresponding M2 profile, $M2_{sim}$. As an M2 profile is the intensity weighted sum of the $\sigma$ values along a line-of-sight, M2 is always an overestimate to the intrinsic $\sigma$. Hence, $M2_{sim}$ will be higher than $M2_{obs}$.  We then calculate the difference between M2$_{obs}$ and M2$_{sim}$ and introduce a proportional change in the $\sigma_{in}$ for the next iteration. Thus, iteratively we obtain a such $\sigma_{in}$ which produces an $M2_{sim}$ reasonably matching to $M2_{obs}$.

We stop the {\it iterative} method when $M2_{sim}$ matches to the $M2_{obs}$ with better than one percent accuracy. However, this accuracy is calculated over the entire profile excluding the central region of twice the beam size. Due to finite beam size, the M2 values at the central part contains a significant contribution from the $\sigma$ values within a region of $\sim$ a beam size around $R=0$. As a result, a higher $\sigma$ values at these areas saturates the M2 values (i.e., doesn't allow it to decrease) at the centre which in turn saturates the convergence accuracy. 

We evaluate the performance of the {\it iterative} method by applying it to simulated clouds for which the structural parameters are already known to us. As we want to apply this method to UCHVCs/CHVCs or potential dark galaxies, we simulate two types of \HI~clouds, one without any dark matter and the other with dark matter. We assume a cloud size of 300 pc for a 3 sigma \HI~sensitivity of $0.12~M_{\odot} pc^{-2}$ ($1.5 \times 10^{19}$ \acc) and a linear intrinsic $\sigma$ profile of the form $\sigma (R) = 0.012\times R + 4$, where $R$ is in parsec and $\sigma$ is in km/s. We used the observed beam size to be $15^{\prime \prime} \times 15^{\prime \prime}$. The input parameters are motivated by the observation of Leo T and typical to UCHVCs/CHVCs. With these inputs and known $\sigma$ profile, we simulate the two \HI~clouds by solving Eq.~\ref{eq4} and consequently produce their \HI~moment profiles. We then use these moment profiles as the inputs to the {\it iterative} method to evaluate its performance.

\begin{figure}
\begin{center}
\begin{tabular}{c}
\resizebox{0.45\textwidth}{!}{\includegraphics{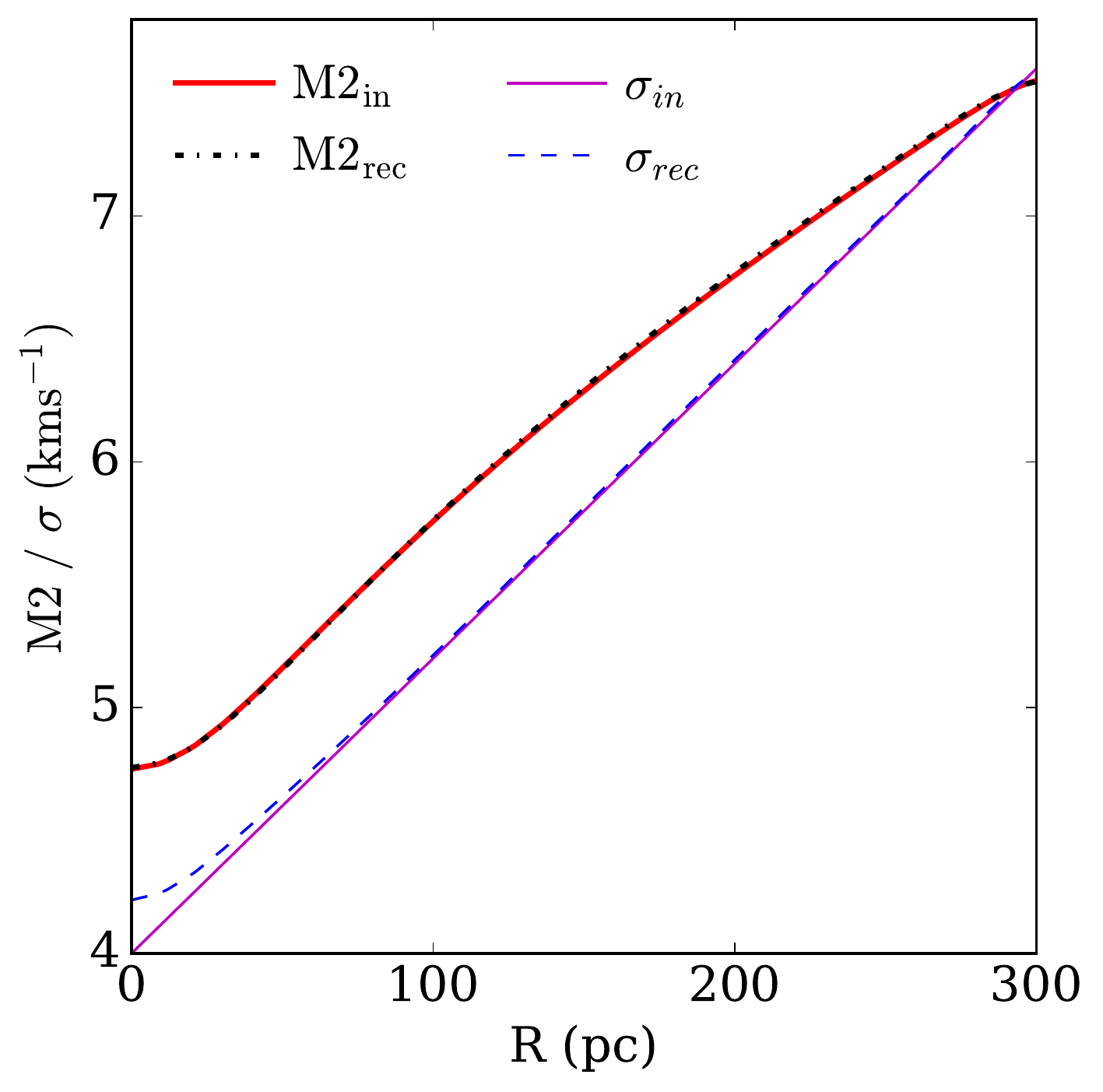}} \\
%\resizebox{55mm}{!}{\includegraphics{figures/ndm_rho_1pc.pdf}} &
\resizebox{0.45\textwidth}{!}{\includegraphics{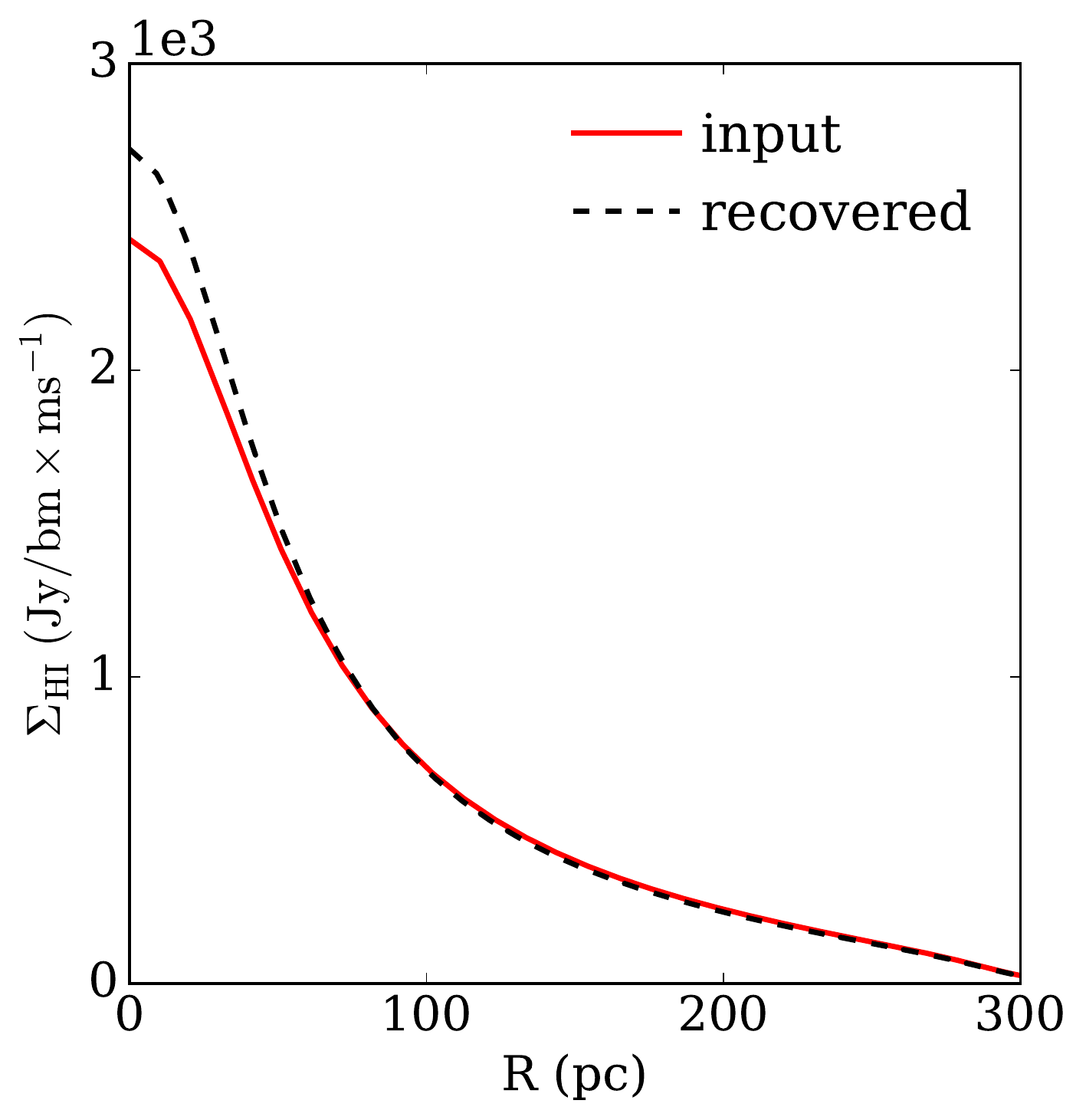}} 
\end{tabular}
\end{center}
\caption{The results of the {\it iterative} method in the absence of any dark matter halo. Top panel: Show the recovery of the M2 profile by the {\it iterative} method. The red and the solid magenta lines represent the input M2 and the $\sigma$ profiles respectively. Whereas the dot-dashed black and the dashed blue lines represent the recovered M2 and the $\sigma$ profiles respectively. Bottom panel: The corresponding input (solid red line) and the recovered (black dashed line) \shi~profiles.}
\label{itr}
\end{figure}

In Fig.~\ref{itr} we show the performance of the {\it iterative} method on an \HI~cloud without any dark matter in it. The top panel shows the input and the recovered M2/$\sigma$ profiles whereas the bottom panel shows the input (solid red curve) and the retrieved (black dashed curve) $\Sigma_{HI}$ profiles. As can be seen from the figure, the {\it iterative} method could, in fact, estimate a reasonable $\sigma$ profile which can produce other observables with acceptable accuracy. The estimated total intensity profile by the {\it iterative} method at the centre deviates from the actual one by $\lesssim ~10-15\%$ due to convergence stiffness at the centre. However, at the outer radii, the recovered and the assumed $\Sigma_{HI}$ profile matches well within a few percent.

\begin{figure}
\begin{center}
\begin{tabular}{c}
\resizebox{0.45\textwidth}{!}{\includegraphics{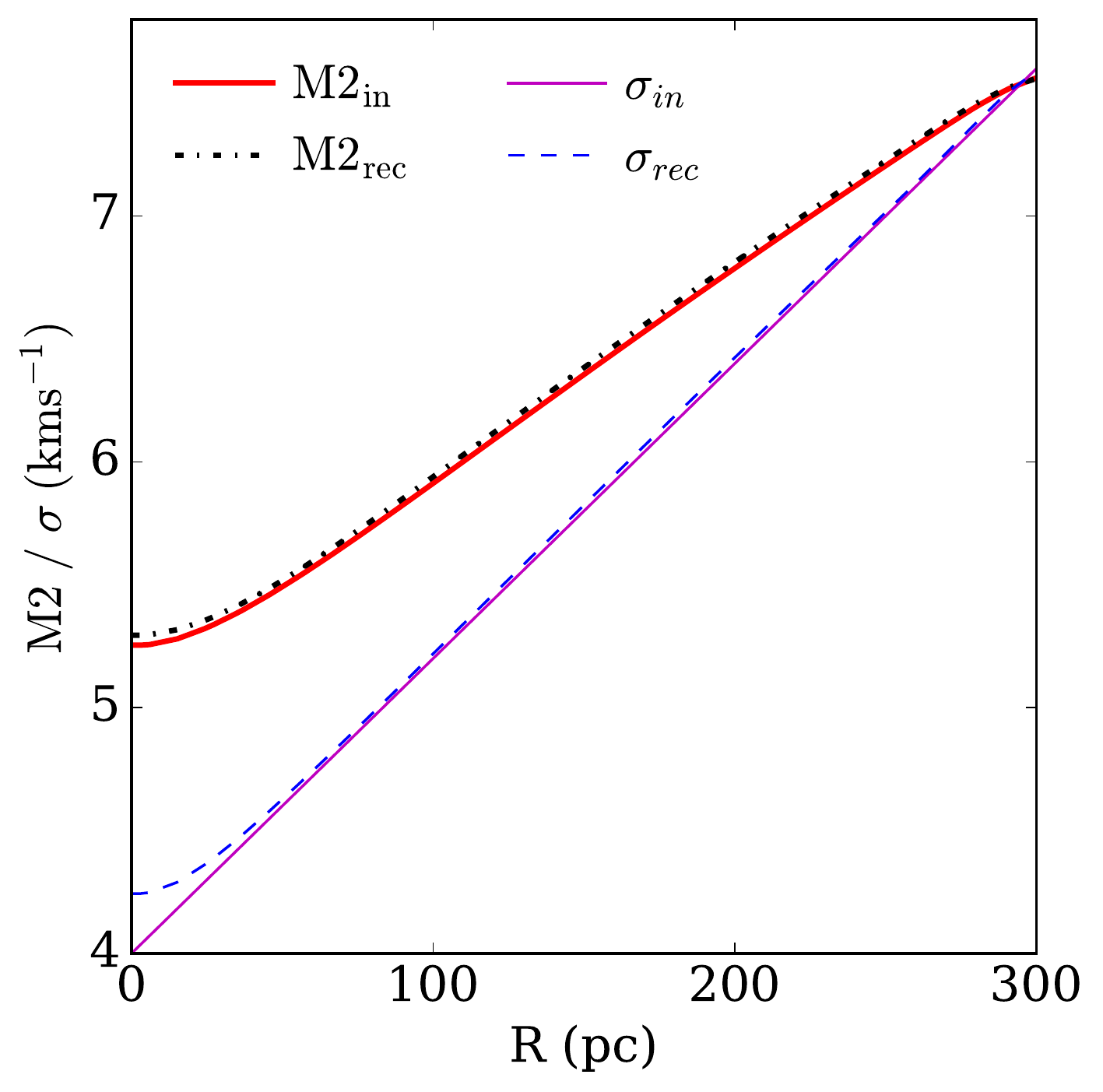}} \\
%\resizebox{58mm}{!}{\includegraphics{figures/dm_iso_rho.pdf}} \\
\resizebox{0.45\textwidth}{!}{\includegraphics{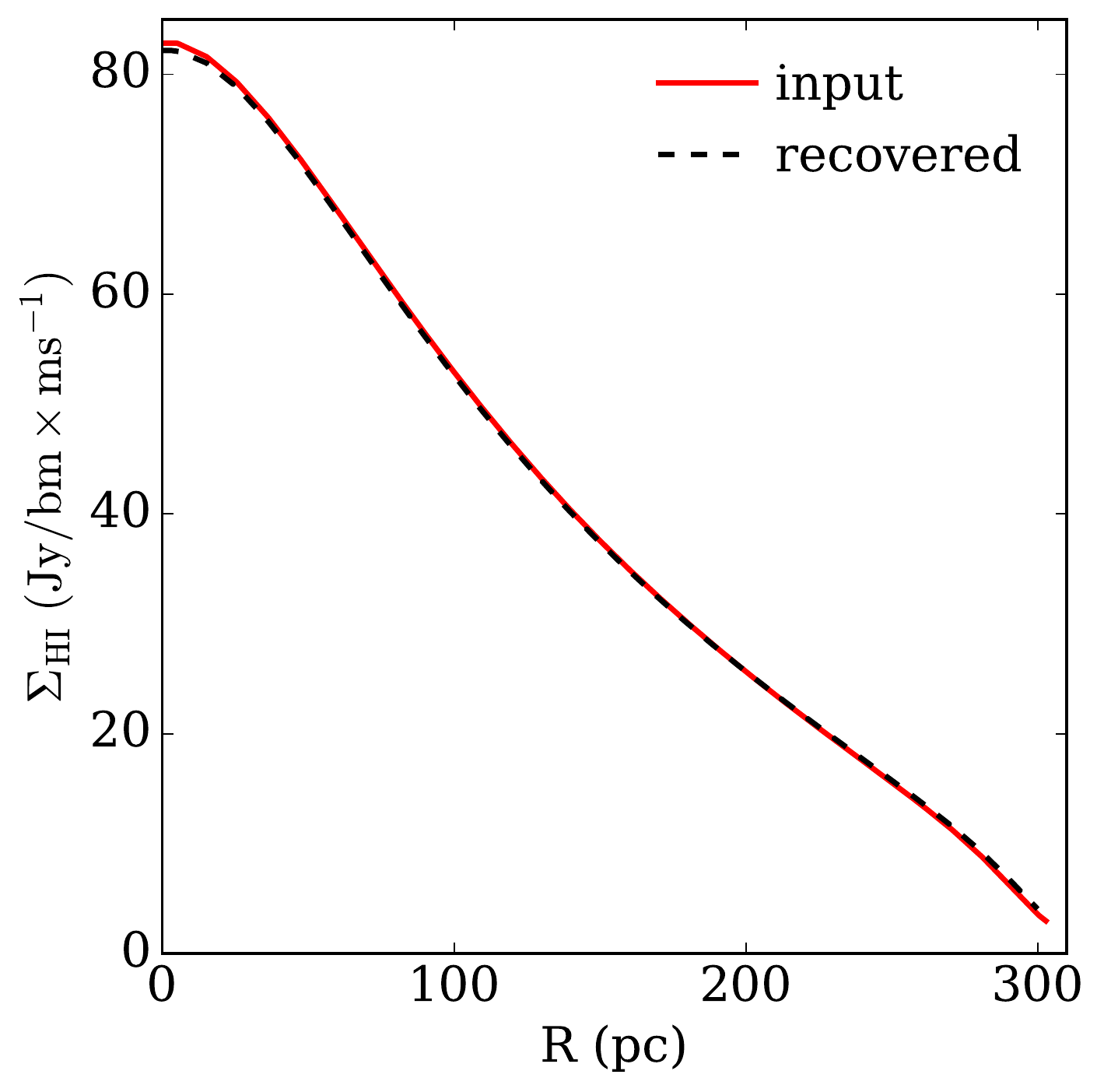}} 
\end{tabular}
\end{center}
\caption{The results of the {\it iterative} method with the presence of an isothermal dark matter halo. The legends in both the panels are exactly the same as in Fig.~\ref{itr}. See text for more details.}
\label{iso}
\end{figure}

\begin{figure*}
\begin{center}
\begin{tabular}{c}
\resizebox{\textwidth}{!}{\includegraphics{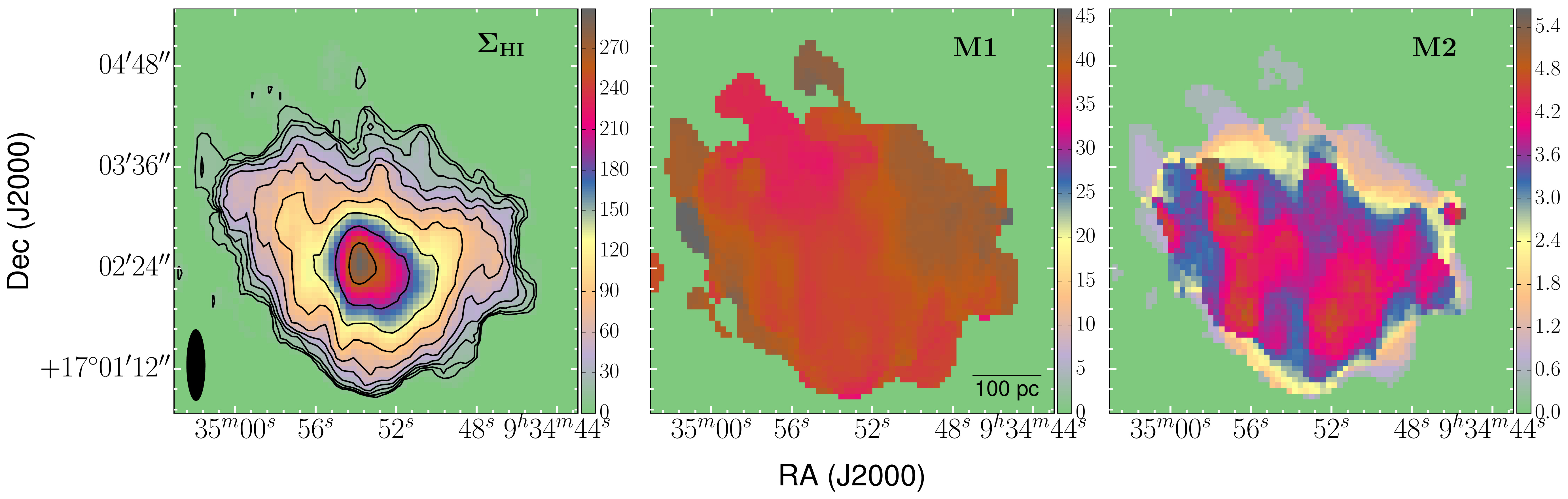}}
\end{tabular}
\end{center}
\caption{The \HI~moment maps of Leo T. The left panel shows the \HI~total intensity distribution in color scale along with the contours. The contour levels are $(2, 2.8, 4, 5.6, .....) \times  10^{19} $~\acc. The middle panel shows the heliocentric velocity field (systemic velocity at 38.6 \kms~\citep{ryan08}) and the right panel shows the M2 distribution.  The color-bars in the middle and right panels are in the units of \kms. The beam is shown at the bottom left corner of the left panel.}
\label{mom}
\end{figure*}

Next, we apply the {\it iterative} method to the simulated \HI~cloud which is hosted by a dark matter halo. We assumed the same cloud parameters as used above. Motivated by the observed dark matter density distribution in nearby dwarf galaxies \citep{moore94,deblok96,deblok97,deblok01,deblok02,weldrake03,spekkens05,
kuziode06,kuziode08,oh11a,oh11b}, we use a pseudo-isothermal dark matter density profile for our simulated cloud. A pseudo-isothermal profile can be given by
\begin{equation}
\label{eq_iso}
\rho(R) = \frac {\rho_0}{1 + \left(\frac{R}{r_s}\right)^2}
\end{equation}

\noindent where the characteristic density $\rho_0$ is the central dark matter density at $R=0$ and $r_s$ is the scale radius. These two parameters completely describe a spherically symmetric pseudo-isothermal dark matter halo. For this particular case we used $\rho_0 = 0.1$ \mspcc~and $r_s = 100$ pc.

In figure~\ref{iso} we show the results of the {\it iterative} method for the \HI~cloud with dark matter halo. It can be seen from the figure, the recovered moment profiles well compare with the input ones, and in all cases, the discrepancy contained within a few percent. This, in turn, implies that if one can adequately guess the dark matter halo parameters, the M2 and the $\Sigma_{HI}$ profile can be recovered with reasonable confidence by the {\it iterative} method. However, there might be degeneracies in the dark matter halo parameters which we discuss in details in the context of Leo T. In the following section; we apply this {\it iterative} method to the observed data of Leo T to understand its hydrostatic structure and attempt a mass-modelling.

\section{Results and discussion}
\label{application}

Leo T was discovered in SDSS data release 5 (DDR5) by \citet{irwin07}. With an estimated distance of $\sim$ 420 kpc, the \HI~size and the \HI~mass of this galaxy are found to be $\sim$ 300 pc and $\sim 2.8 \times 10^5$ \ms~\citep{ryan08} respectively. It has a very faint stellar population amounting to a stellar mass of $\sim 10^5$ \ms \citep{irwin07}. Though no signature of rotation is found in the \HI~velocity field of this galaxy, the presence of a faint stellar population confirms it to be one of the faintest galaxy known. The \HI~properties of this object is highly similar to that of the UCHVCs/CHVCs, and hence it is an ideal candidate to apply our hydrostatic modelling.

In Fig.~\ref{mom} we plot \HI~moment maps of Leo T as observed by ~\citet{ryan08} using the Westerbork Synthesis Radio Telescope (WSRT). The left panel shows the \shi~map with contour levels starting from $2 \times 10^{19}$ \acc and separated by $\sqrt{2}$. The middle panel shows the \HI~velocity field as traced by the first moment of the \HI~spectral cube and the M2 map is shown in the right panel. All the images are at a spatial resolution of $\sim 13$\arcs$\times 50$\arcs~which translates to a linear scale of $\sim$ 25$\times$100 pc at the distance of Leo T. From the left panel, it can be seen that the \shi~distribution of Leo T looks reasonably symmetric. Not only that, the velocity field also looks uniform without any sign of a systematic velocity gradient which enforces the idea that Leo T is a thermally supported system under hydrostatic equilibrium.  

\begin{figure}
\begin{center}
\begin{tabular}{c}
\resizebox{0.48\textwidth}{!}{\includegraphics{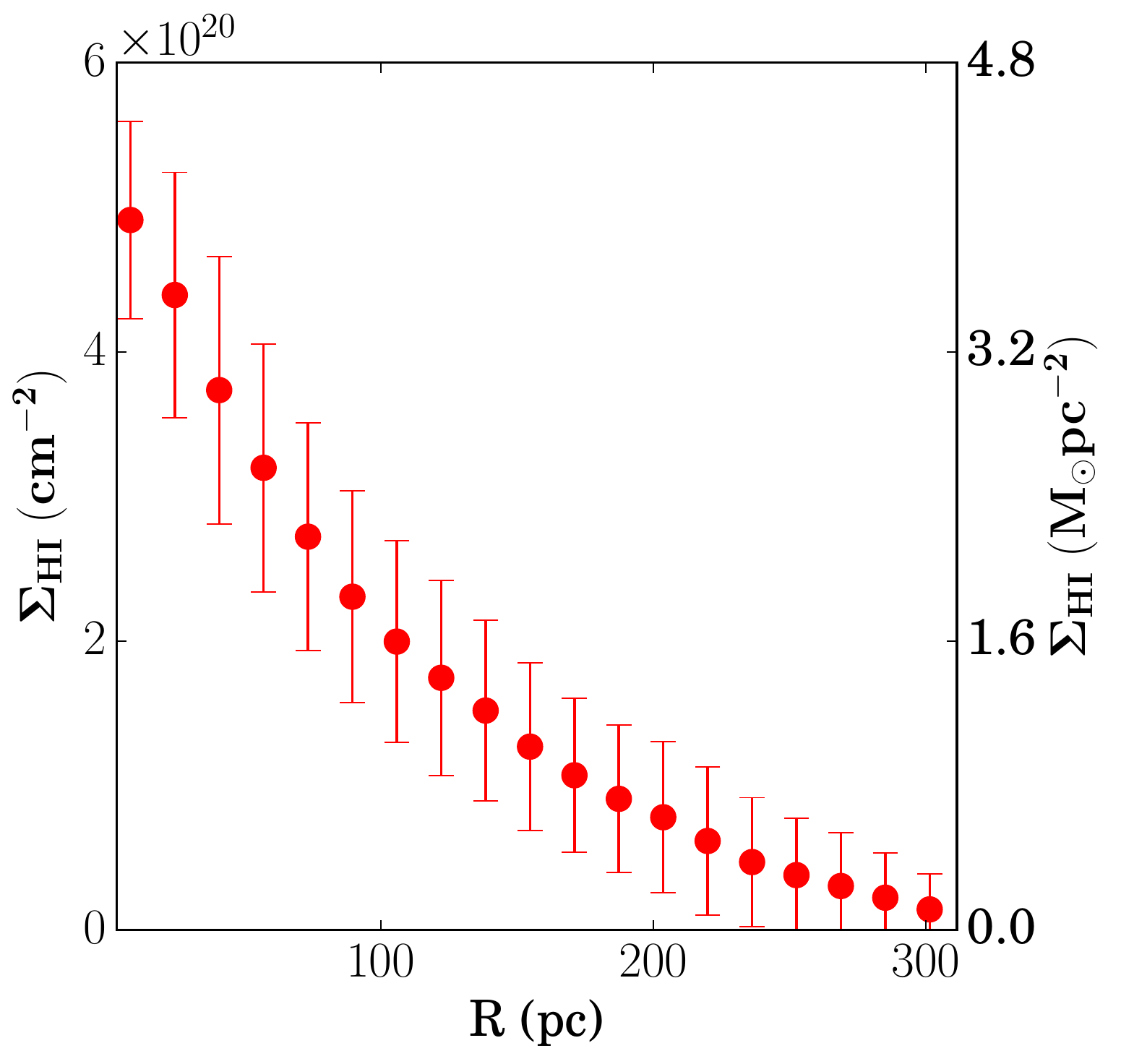}} \\
\resizebox{0.43\textwidth}{!}{\includegraphics{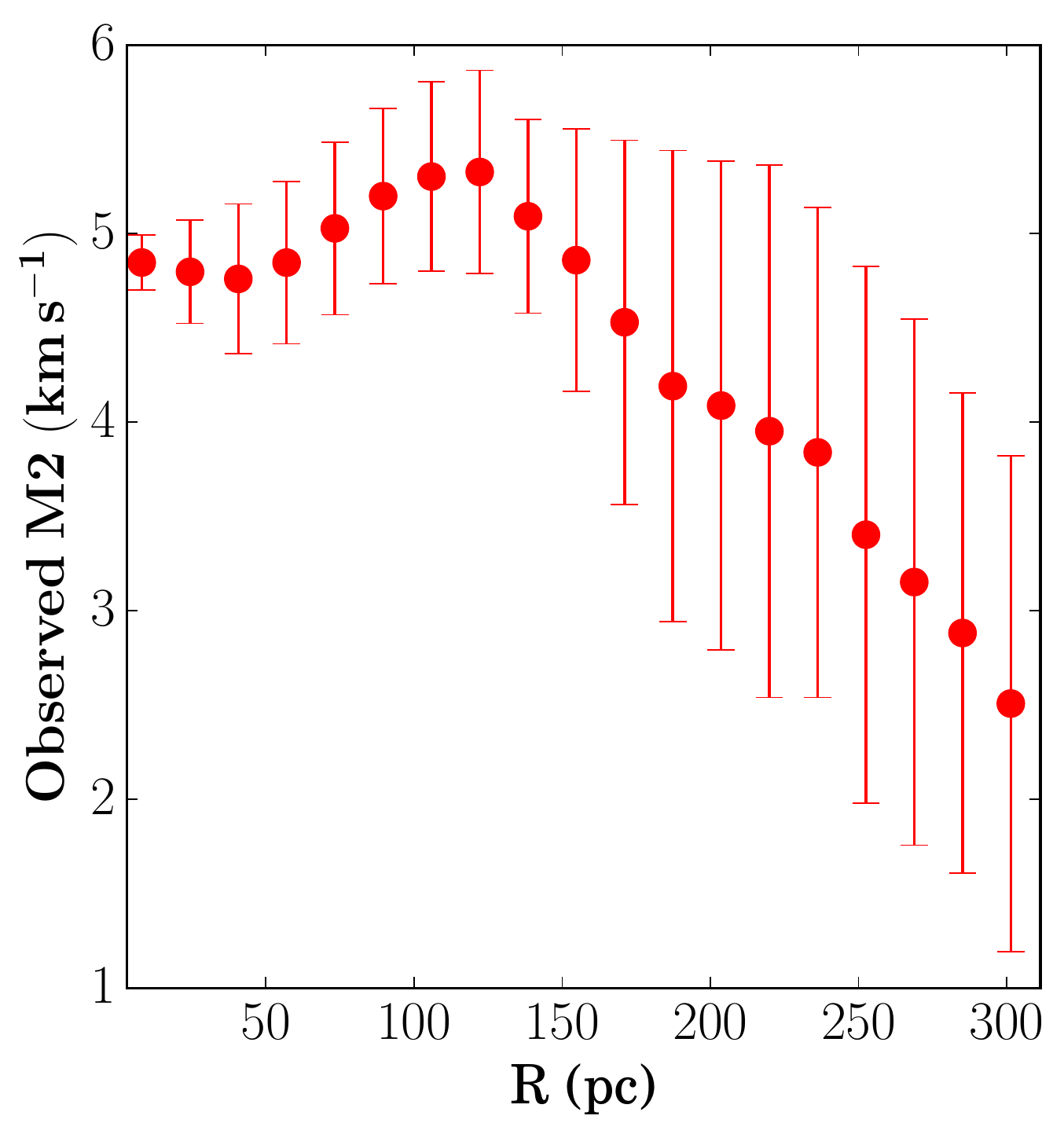}} 
\end{tabular}
\end{center}
\caption{Top panel: The azimuthally averaged observed radial \shi~profile of Leo T. Bottom panel: Corresponding radial M2 profile.}
\label{mom_prof}
\end{figure}

To perform a self-consistent mass modelling of Leo T, one needs to calculate the \shi~and the M2 profiles. In Fig.~\ref{mom_prof} top and bottom panels we plot the radial profiles of \shi~and M2 respectively. The radial profiles are calculated by azimuthally averaging all the values within a radial bin. To calculate the error bars in the \shi~profile, we first determine the uncertainties in the estimation of the zeroth moment from the spectral cube. This uncertainty is then added in quadrature to the statistical variation of \shi~values within a radial bin to calculate the final error bars. As can be seen from the top panel of the figure, the \shi~profile extends roughly up to 300 pc which is assumed to be the extent of the \HI~distribution in Leo T. From the bottom panel, one can see that the M2 profile decreases as a function of radius after an initial increase at $R \lesssim$ 50 pc. The error bars on the M2 profile are calculated using the same technique as it was used to calculate the error bars on the \shi~profile. However, M2 measurements are known to be contaminated by low SNR spectra in the data cube, especially at the outer regions. As a result, the M2 at these areas only trace the peaks of the individual \HI~spectrum, which in turn artificially decreases the velocity width leading to a misinterpretation of the data.

To overcome this SNR problem, we adopt a spectral stacking technique. Given the symmetry in the \shi~map, the \HI~spectra within a radial bin are assumed to be similar and stacked after aligning their central velocities. To align the individual spectrum in velocity, one needs to estimate the central velocities. We adopt a similar approach as used by \citet{deblok08} to estimate the central velocity of an \HI~spectrum. We fit line-of-sight \HI~spectra with SNR $\gtrsim 5$ with a Gaussian-Hermite polynomial to estimate its central velocity. All spectra within a radial bin are then stacked together to generate a high SNR spectrum. These high SNR stacked spectra are then used to create intensity weighted second moment or M2 profile which now have much higher SNR than individual spectrum in the data cube.

\begin{figure}
\begin{center}
\begin{tabular}{c}
\resizebox{0.45\textwidth}{!}{\includegraphics{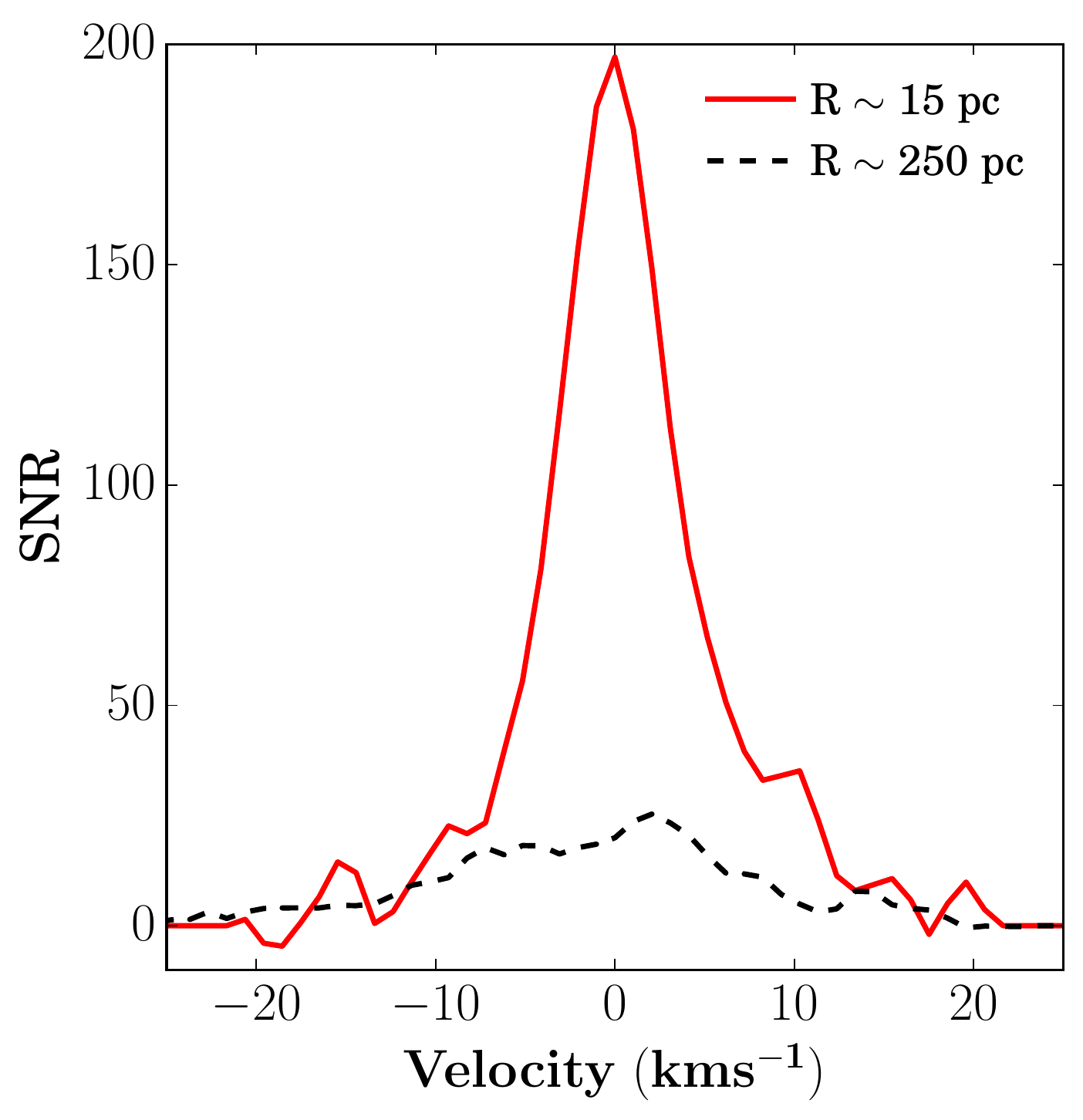}}
\end{tabular}
\end{center}
\caption{Stacked \HI~profiles at two extreme radial bins of Leo T. The solid red line represents the stacked \HI~spectrum at the first radial bin whereas the black dashed line represents the same at the last radial bin.}
\label{stacked_prof}
\end{figure}

It should be noted that the individual \HI~spectrum must have a minimum SNR of 5 to qualify for fitting with a Gaussian-Hermite polynomial. This criterion restricts the radial extent till which one can generate an M2 profile from stacked spectra. We have used low resolution (30 \arcs) data cube to achieve a higher SNR to larger radii as compared to high-resolution data. We note that, as there is no rotation observed in Leo T, spectral broadening due to rotation within a beam would be negligible. To sample the radial M2 profile sufficiently, we choose the radial bin to be half of the beam width. With this criteria, we could generate M2 profile up to $\sim$ 250 pc with a minimum SNR of $\sim$ 25 of the stacked spectra at the furthest radial bin. In Fig.~\ref{stacked_prof} we plot the stacked \HI~spectra in the two extreme radial bins. As can be seen, the SNRs of these stacked spectra are significantly higher than the maximum SNR of individual spectra, which is $\sim$ 5 at the furthest radial bin.

\begin{figure}
\begin{center}
\resizebox{0.45\textwidth}{!}{\includegraphics{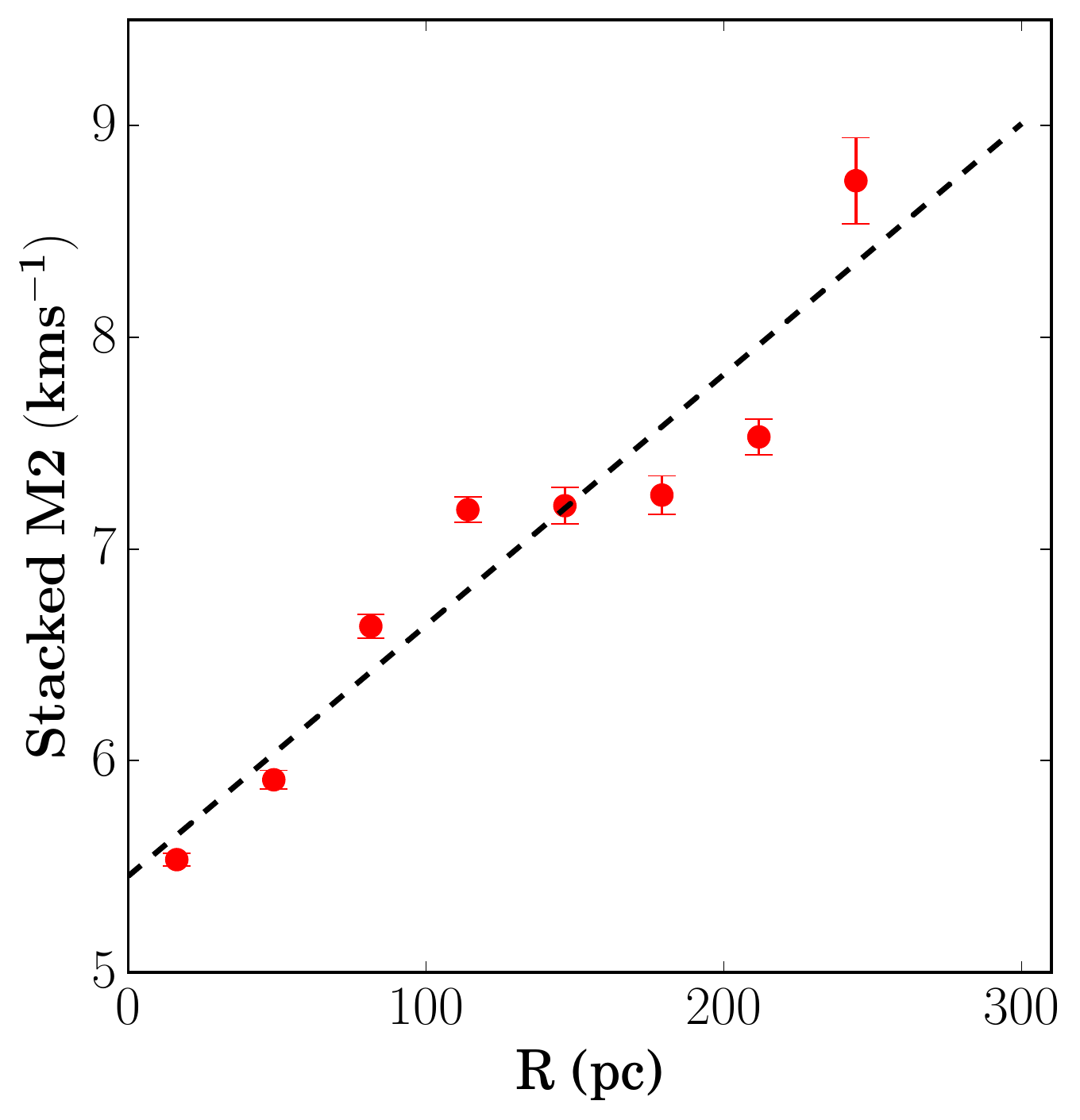}}
\end{center}
\caption{The M2 profile of Leo T as extracted from the stacked spectra. The black dashed line represents a linear fit to the profile.}
\label{m2prof}
\end{figure}

In Fig.~\ref{m2prof} we plot the corresponding M2 profile. The error bars in the figure are estimated by accounting the noise in the stacked spectra (same way as done in Fig.~\ref{mom_prof} except excluding the statistical variation). The reduced error bars on the new M2 profile is thus an outcome of the significantly higher SNR of the stacked spectra. We note that the hydrostatic equation (Eq.~\ref{eq4}) has terms with second derivatives in thermal width ($\sigma$). However, given the sensitivity of the observation, we do not expect to capture a second-order variation in the M2 profile and assume that a linear fit is a reasonable approximation of the M2 profile. Nonetheless, from the figure, it can be seen that a straight line well describes the M2 profile, given the fact that the spectral resolution is 1 \kms. The slope of the fitted M2 profile is  0.012 $\pm$ 0.002 \kms$\rm pc^{-1}$ with an intercept of 5.5 $\pm$ 0.2 \kms.

Next, we apply the {\it iterative} method to Leo T. As Leo T is very similar to UCHVCs/CHVCs, at first, we will assume that there is no dark matter in it. Following the same prescription as described in \S~\ref{itr_method}, we solve Eq.~\ref{eq4} for Leo T.  However, unlike the cloud we solved in \S~\ref{itr_method}, Leo T found to have a stellar component. Though the stellar mass is less than half the gas mass, we include this component in Eq.~\ref{eq4} for consistency. A Plummer profile found to well describe the observed stellar distribution \citep{irwin07} which is given by

\begin{equation}
\rho_s(R) = \left(\frac{3M}{4\pi a^3} \right) \left(1 + \frac{R^2}{a^2}\right)^{-\frac{5}{2}}
\end{equation}

\noindent where $M$ is the total stellar mass ($ 10^5$ \ms) and $a$ is the Plummer radius (130.8 pc) \citep[see,][for more details]{irwin07}. Unlike the \HI, the stellar component is not considered to be live, i.e., its density distribution is not calculated by solving Eq.~\ref{eq4} but taken to be fixed by the Plummer formula. This is justified as an assumed Plummer density profile reproduces the stellar surface density very well. Nonetheless, the stellar mass present in Leo T is less than half the gas mass, and hence, it is not expected to influence the hydrostatic equilibrium considerably. For the \HI~component, we use an observed cloud size of $\sim$ 300 pc. We also scale the \HI~mass by a factor of 1.4 to account for the presence of Helium.

\begin{figure}
\begin{center}
\begin{tabular}{c}
\resizebox{0.45\textwidth}{!}{\includegraphics{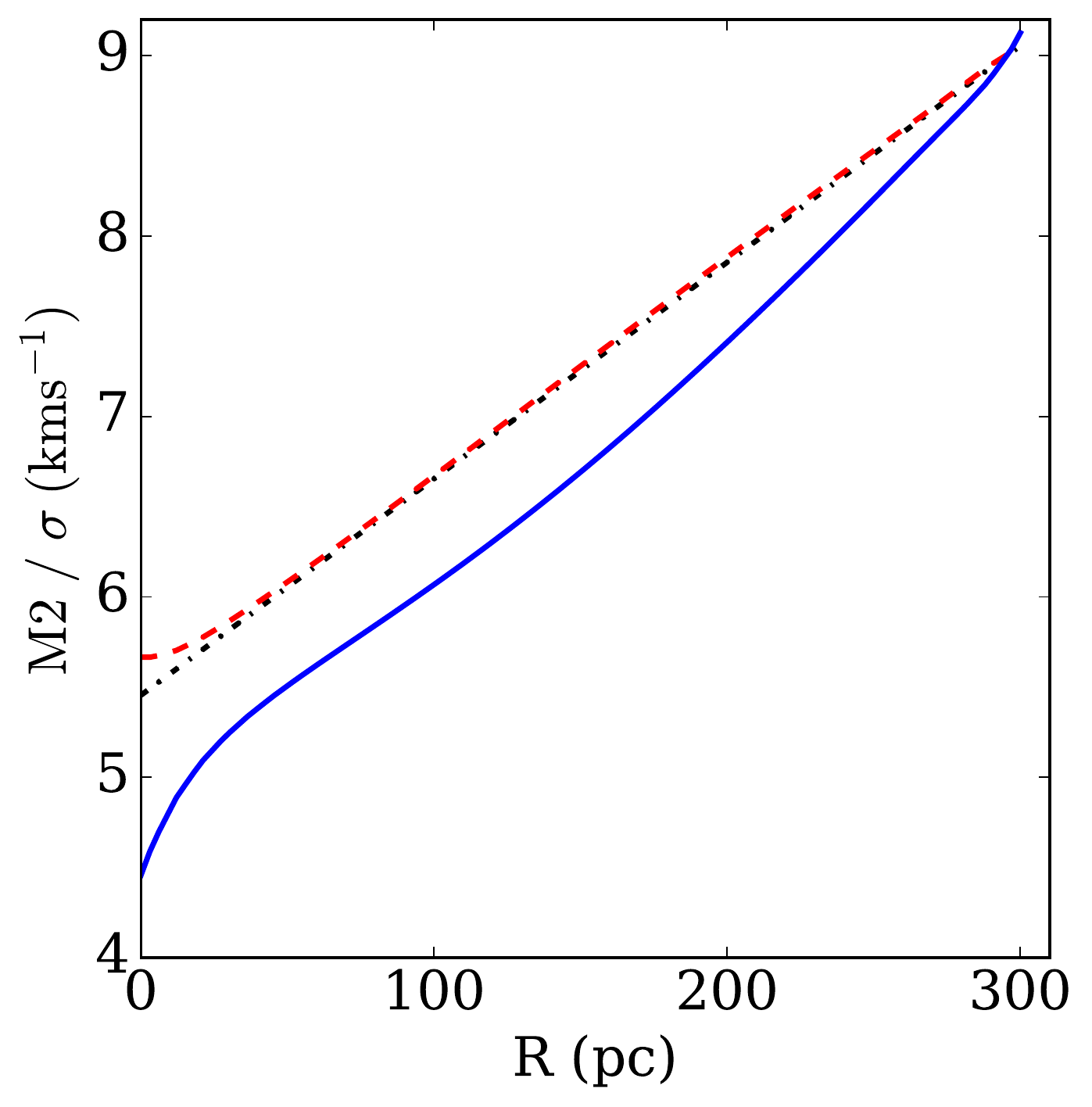}} \\
\resizebox{0.49\textwidth}{!}{\includegraphics{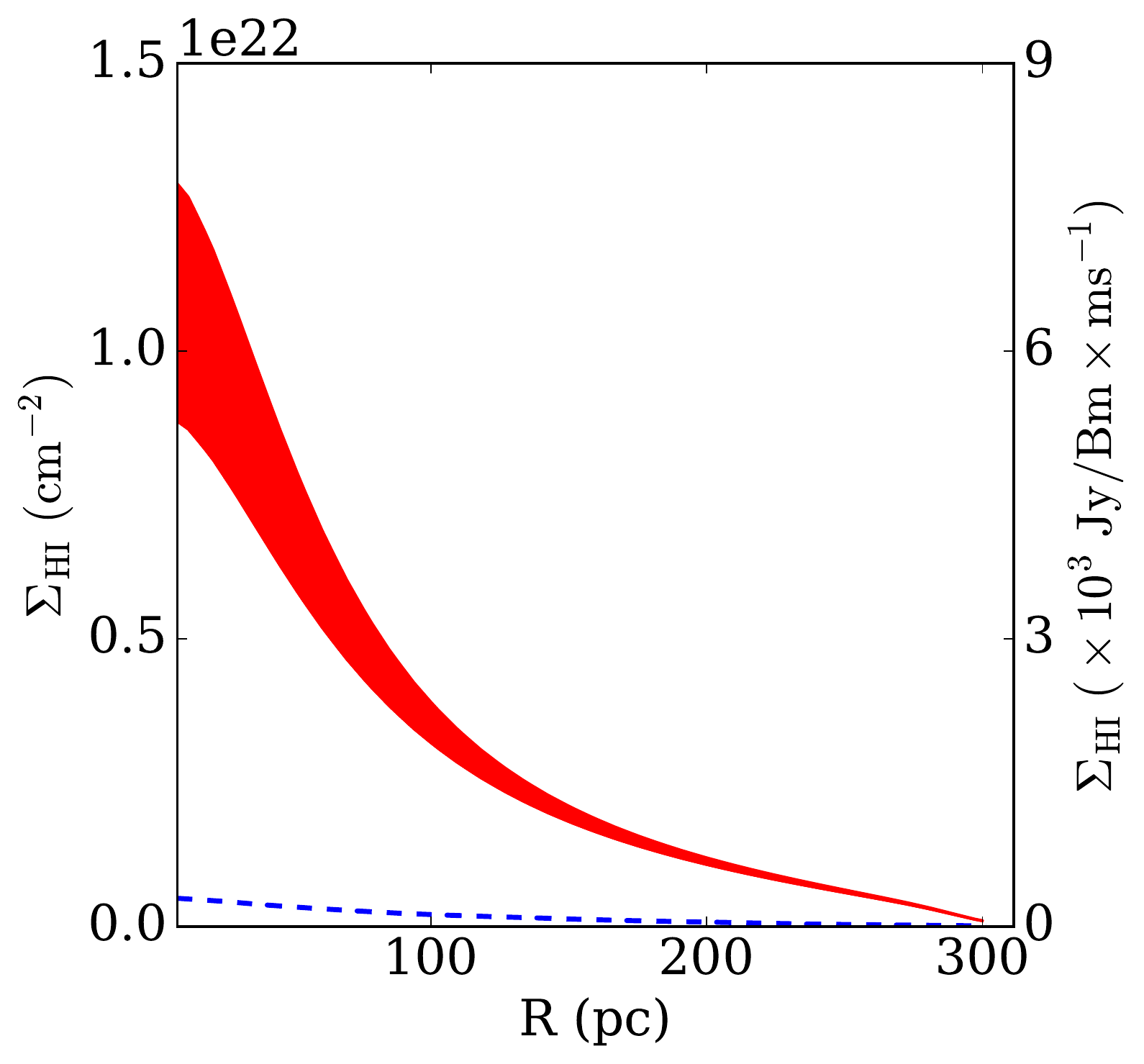}}
\end{tabular}
\end{center}
\caption{Top panel: Shows the observed and the recovered M2 profiles along with the estimated intrinsic velocity dispersion profile for Leo T. The black dashed-dotted line represents the observed M2 profiles whereas the red dashed line represents corresponding recovered M2 profile. The solid blue line represents the estimated intrinsic velocity dispersion profile. Bottom panel: Shows the recovered and the observed \shi~profiles. The red shaded region represents the recovered \shi~profile by the {\it iterative} method whereas the blue dashed line represents the observed one. As can be seen, the recovered \shi~profile falls short to explain the observation if one assumes no dark matter in Leo T. See text for more details.}
\label{ndm_leot}
\end{figure}

In Fig.~\ref{ndm_leot} top panel we show the input and the recovered M2 profiles as well as the corresponding estimated velocity dispersion profiles for Leo T. We note that the recovered M2 profile matches very well with the input M2 profile. In the bottom panel we show the corresponding \shi~profile (red shaded region) as calculated by the {\it iterative} method. This is the required \shi~profile if Leo T has to be in hydrostatic equilibrium {\it without any dark matter}. Corresponding \HI~mass would then be $\sim 10^7$ \ms. Whereas the observed \HI~mass is $\sim 2.8 \times 10^5$ \ms. The blue dashed line in the figure shows the observed \shi~profile. It can be seen that the observed \HI~surface density or mass is much less than what is required for Leo T to be in hydrostatic equilibrium without any dark matter.

%However, it can be shown that, for a self-gravitating virialized isothermal \HI~cloud, the observed column density must be $N_{HI} \gtrsim (3 \sigma^2)/(4 \pi G D \theta m_{HI})$ to be stable under hydrostatic equilibrium. Where, $N_{HI}$ is the average \HI~column density, $\sigma$ is the velocity dispersion, $G$ is the gravitational constant, $D$ is the distance, $\theta$ is the angular size of the cloud and $m_{HI}$ is the mass of the hydrogen atom. For CHVCs/UCHVCs, this formula, in general, is used to check if these clouds are self-gravitating or not. If one puts the parameters of Leo T in the above equation, it can immediately be seen that Leo T cannot be a self-gravitating cloud given its \HI~mass. However, the above formula assumes the cloud to be isothermal, which might not be true in many cases. For example, many CHVCs shows clear $\sigma$ gradient in high resolution \HI~observations \citep{forbq16}. Leo T itself shows a considerable $\sigma$ gradient (see e.g., Fig.~\ref{m2prof}). In these circumstances, one needs to solve the exact equation (i.e., Eq.~\ref{eq4}) to calculate the required \HI~column density or mass accurately.

Next, we try to estimate the dark matter halo parameters for Leo T which can produce the observed \shi~or \HI~mass. To do that, we first construct a large number of dark matter halos with different halo parameters and solve Eq.~\ref{eq4} for each of them to calculate the expected \shi~profiles. We then compare these surface density profiles to the observed one to identify the best matched dark matter halo parameters. 

We choose observationally motivated pseudo-isothermal dark matter halo density profiles to construct our trial dark matter halos. A pseudo-isothermal halo can be described by two halo parameters, i.e., the central density, $\rho_0$ and the scale radius $r_s$ (Eq.~\ref{eq_iso}). We explore the two-dimensional parameter space in $\rho_0$ and $r_s$ to identify the best dark matter halo parameters for Leo T. 

We chose the ranges of $\rho_0$ and $r_s$ in a trial and error basis so that the modelled \shi~profiles enclose the observed one. We vary $\rho_0$ between $0.01-2.0 \ M_{\odot}/pc^{3}$ and $r_s$ within $20-150$ pc in a grid of 100 $\times$ 100 bins. We note that though these limits well comply with the observational or numerical findings \citep{strigari08}, a different set of values for $\rho_0$ and $r_s$ might also produce similar results. We then work out the {\it iterative} method for Leo T in every bin. Solving Eq.~\ref{eq4} with dark matter is a compute-intensive step. However, as the halo parameters are independent, Eq.~\ref{eq4} can be solved in parallel for different dark matter halos. We implement this using MPI based parallel code and run on a High-Performance Computing cluster to speed up the calculations.

\begin{figure}
\begin{center}
\begin{tabular}{c}
\resizebox{0.45\textwidth}{!}{\includegraphics{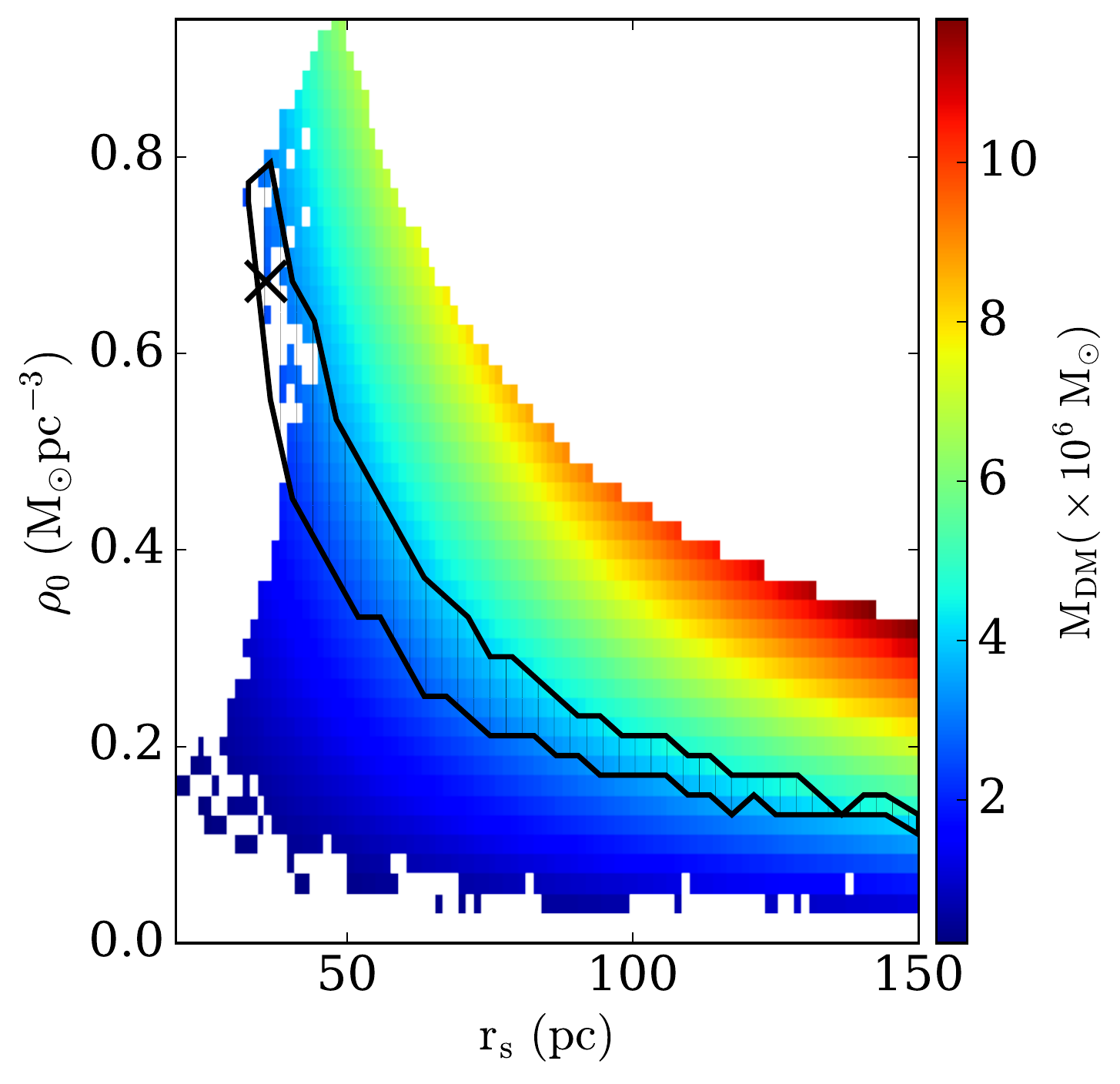}} \\
%\resizebox{55mm}{!}{\includegraphics{figures/rrdm_leot7_max_err.pdf}} &
\resizebox{0.45\textwidth}{!}{\includegraphics{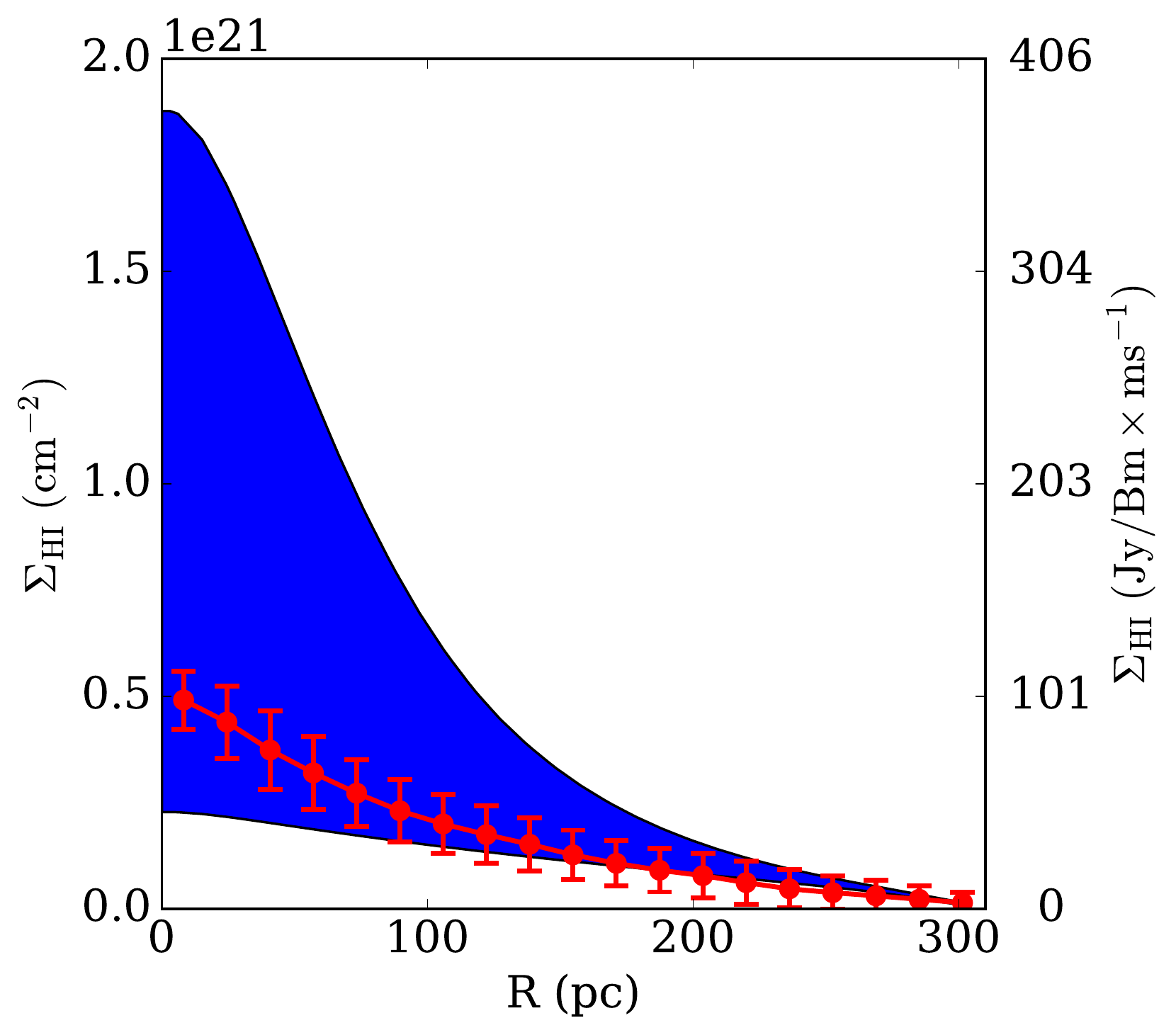}} \\
\end{tabular}
\end{center}
\caption{Top panel: The $\rho_0-r_s$ space for which Eq.~\ref{eq4} is solved. Each pixel in this panel represents an assumed dark matter halo. The colour scale indicates the mass of the dark matter halos (within central 300 pc). The thick black cross with $\rho_0 = 0.67$ \mspcc~and $r_s = 37$ pc indicates the dark matter halo which produces the best-matched \shi~profile to the observation. The hatched region includes all the dark matter halos which produce \shi~profiles consistent with the observed one within the observational uncertainties. Bottom panel: The blue shaded region confines the \shi~profiles for all the dark matter halos as indicated in the top panel. The solid red circles with error bars represent the observed \shi~profile.}
\label{params_leot}
\end{figure}

\begin{figure}
\begin{center}
\begin{tabular}{c}
\resizebox{80mm}{!}{\includegraphics{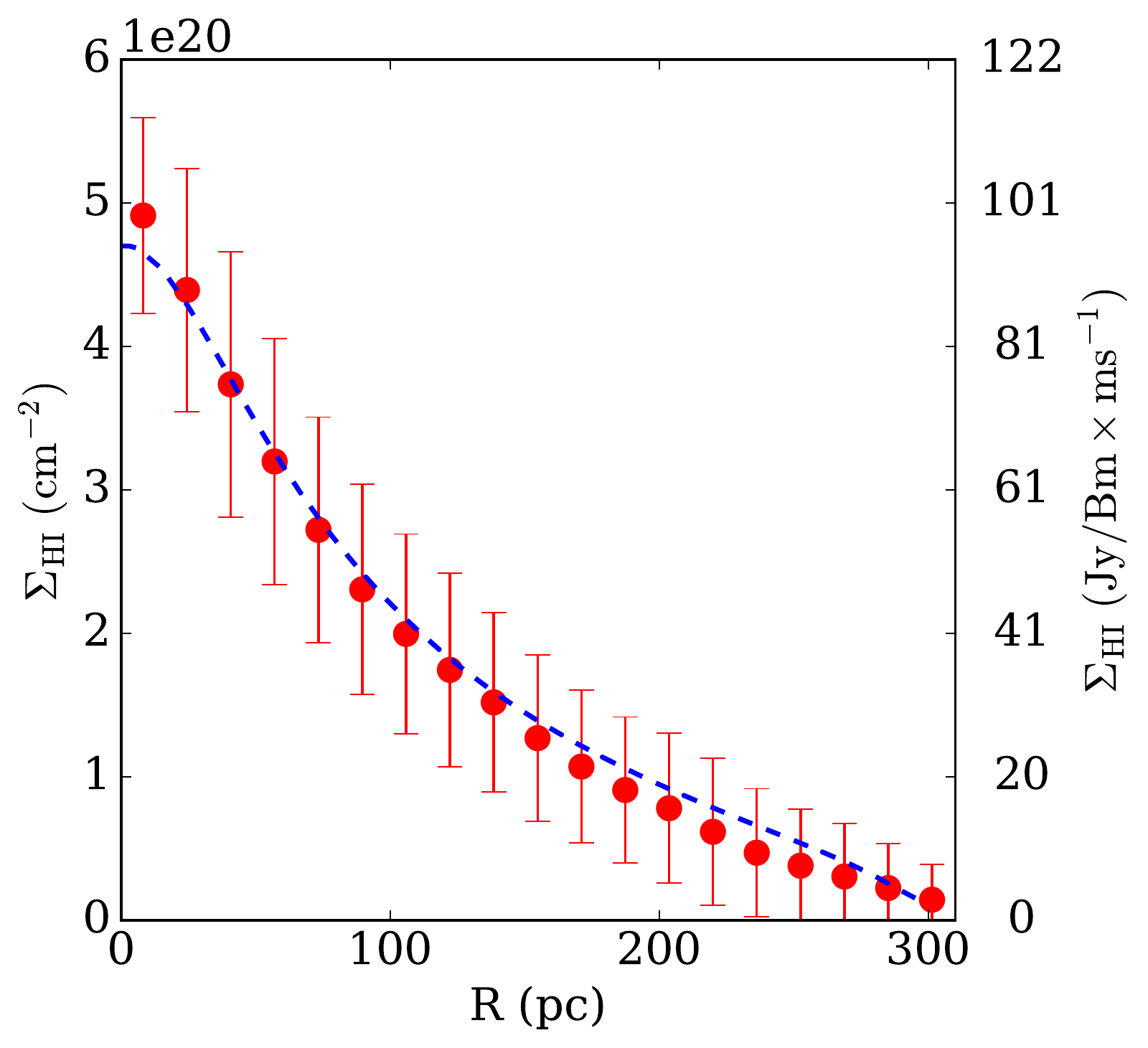}} \\
\resizebox{80mm}{!}{\includegraphics{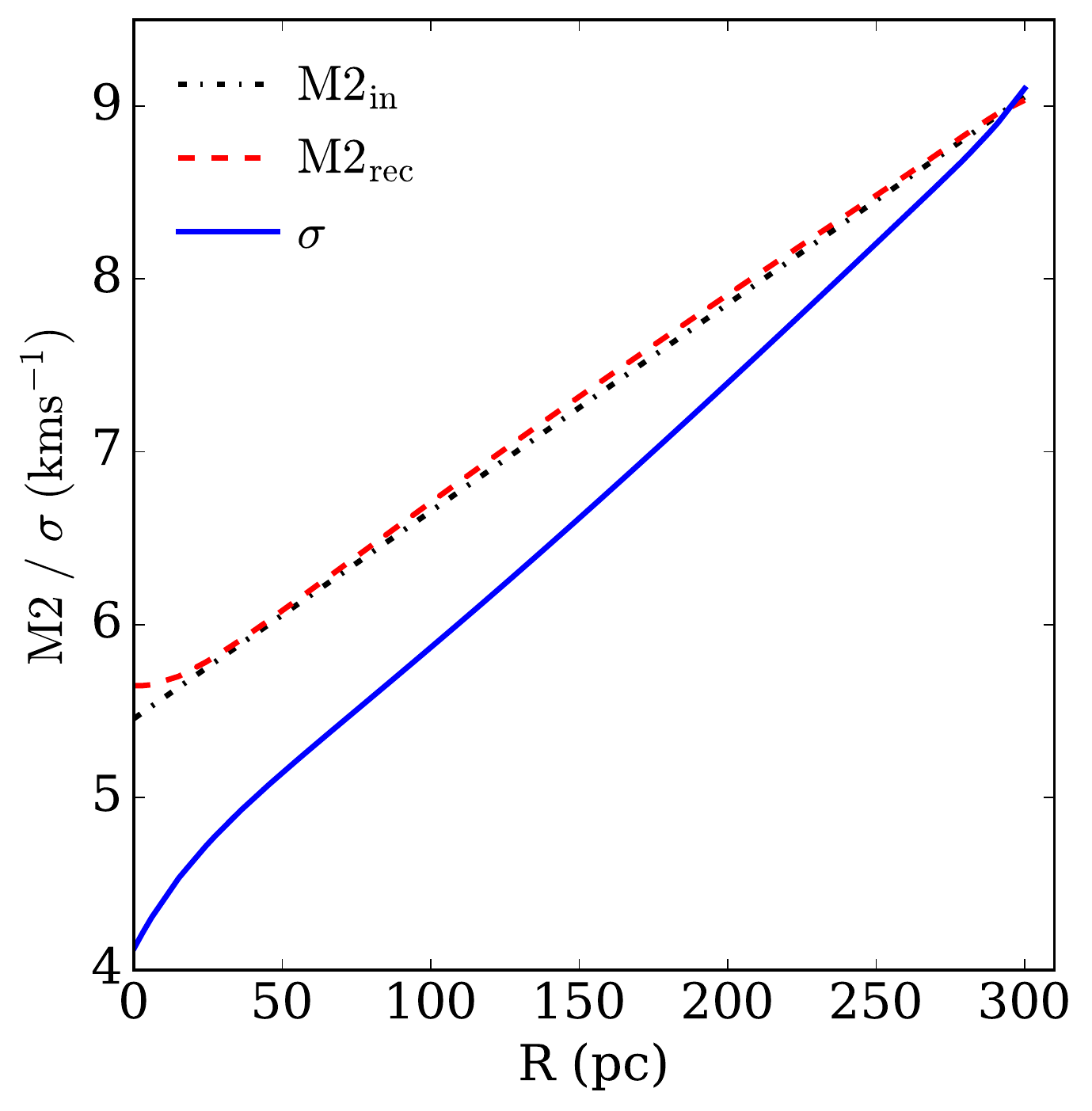}} 
%\resizebox{53mm}{!}{\includegraphics{figures/rrdm_leot7_rho.pdf}} \\
\end{tabular}
\end{center}
\caption{ Top panel: Shows the best-matched \shi~profile for Leo T as recovered by the {\it iterative} method. The solid red circles with error bars represent the observed \shi, whereas the blue dashed line represents the best-matched \shi~profile produced by a dark matter halo with $\rho_0 \simeq 0.67$ \mspcc~and $r_s \simeq 37$ pc. Bottom panel: Shows the corresponding M2 and $\sigma$ profiles. The black dashed-dotted and the red dashed lines represent the observed and the recovered M2 profiles respectively whereas the solid blue line represents the estimated $\sigma$ profile. As can be seen from the top panel, the {\it iterative} method recovers a dark matter halo which reasonably well produces the observed \shi~profile.}
\label{best_prof}
\end{figure}

In Fig.~\ref{params_leot} top panel, we show the parameter space of $\rho_0$ and $r_s$ for which we solve Eq.~\ref{eq4}. The colour scale indicates the mass of the dark matter halos within their central 300 pc region. The choice of the parameter space explores a wide range of dark matter halo mass from $1.4 \times 10^4$ \ms~to $7.6 \times 10^7$ \ms. However, all of these dark matter halos do not produce a hydrostatically stable structure of size 300 pc. For example, a highly massive dark matter halo with large gravity would produce a smaller cloud for a fixed pressure profile (decided mostly by the intrinsic $\sigma$ profile). On the other hand, a very light dark matter halo would need a large amount of \HI~to produce the extra amount of gravity required to make the cloud stable. This, in turn, would result in a large \HI~column density. We exclude both these types of dark matter halos on a trial and error basis as they are not useful to produce the observed \shi. However, we make sure that the rest of the dark matter halos produce \shi~profiles which enclose the observed one very well. In the bottom panel of Fig.~\ref{params_leot}, the blue shaded region confines all such \shi~profiles produced by the valid halos whereas the red circles with error bars represent the observed \shi~profile. 

We use a $\chi^2$ minimization method to identify the model \shi~profile which matches best to the observed one. In Fig.~\ref{best_prof} top panel, we show the best-matched profile (blue dashed line) along with the observed one (solid red circles with error bars). The observed and the recovered M2 profiles along with the corresponding $\sigma$ profile are shown in the bottom panel. It can be noted that the retrieved M2 profile (red dashed line) well matches with the input one (black dashed-dotted line) with a convergence accuracy better than $1\%$. This indicates a reasonably well estimation of the $\sigma$ profile (solid blue line) by the {\it iterative} method. As it can be seen from the top panel of Fig.~\ref{best_prof}, the observed and the best matched $\Sigma_{HI}$ profiles compare very well with each other. The best matched dark matter halo parameters are found to be $\rho_0 \simeq 0.67$ \mspcc and $r_s \simeq 37$ pc which is marked by a thick black cross in the top panel of Fig.~\ref{params_leot}. The mass of the dark matter halo within 300 pc is found to be $\sim 2.7 \times 10^6$ \ms. We note that the dynamical mass of Leo T assuming an isothermal constant $\sigma$ of 7 \kms~is $\sim 10^7$ \ms, which is $\sim 3$ times larger than what we estimate here.

\begin{figure}
\begin{center}
%\resizebox{0.45\textwidth}{!}{\includegraphics{figures/rrdm_leot8_m0_prof_error.jpg}}
\resizebox{0.45\textwidth}{!}{\includegraphics{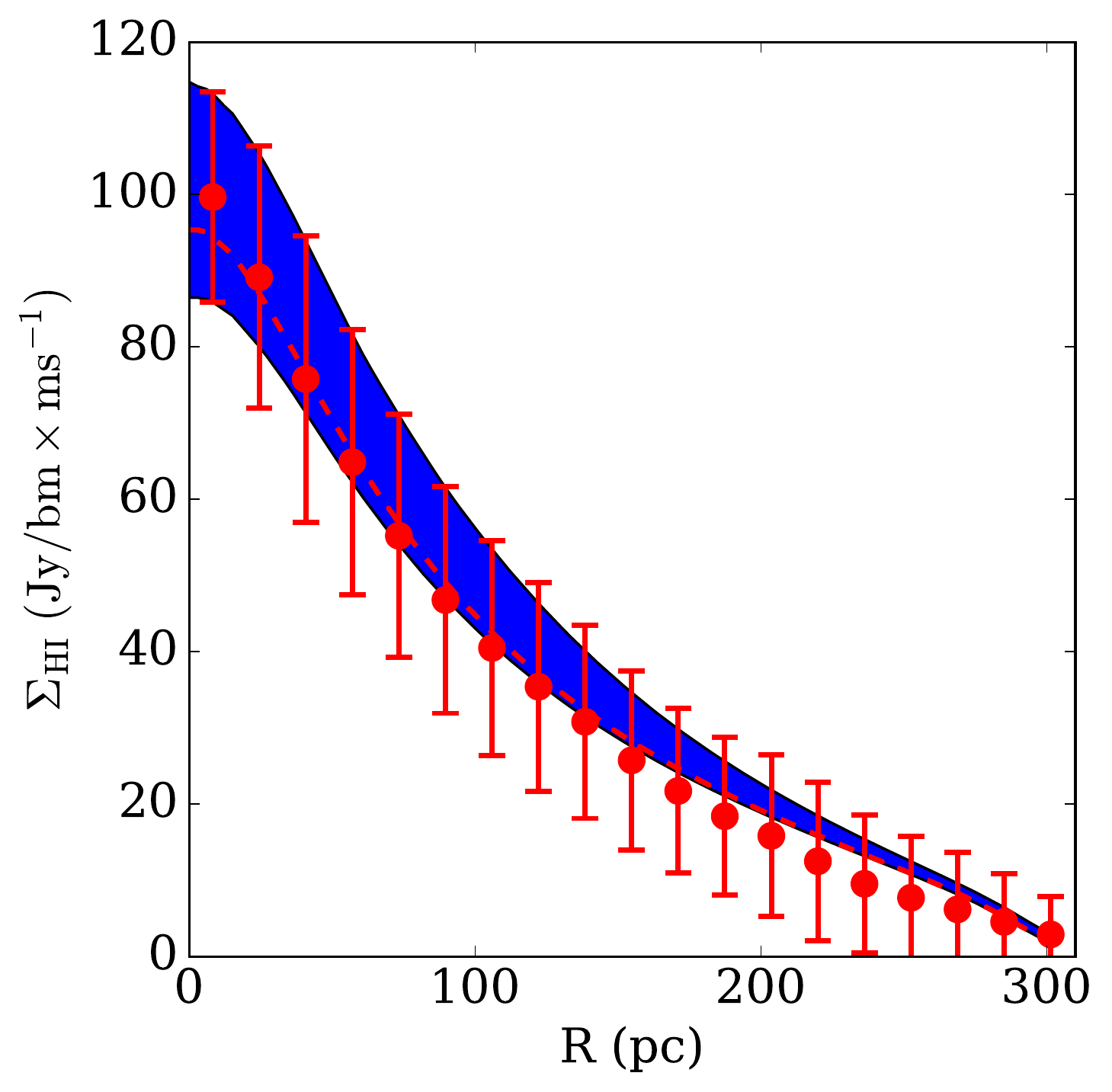}}
\end{center}
\caption{The solid red circles with error bars represent the observed \shi~profile for Leo T whereas the red dashed line represents the best-matched one. The blue shaded region confines all the \shi~profiles which match with the observation within one-sigma error at all radius. The hatched region in Fig.~\ref{params_leot} represents corresponding dark matter halos.}
\label{m0_err}
\end{figure}

Next, we identify all the modelled \shi~profiles which entirely lies within the error bars of the observed profile. We expect that given the uncertainties in the observed $\Sigma_{HI}$ profile, these profiles are reasonably fit to match the observation. In Fig.~\ref{m0_err}, the blue shaded region confines all such profiles whereas the solid red circles with error bars represent the observed \shi~profile. These halos are equally suitable to host Leo T. Interestingly, all of these dark matter halos lie in an envelope of approximately equal dark matter halo mass as shown by the hatched region in Fig.~\ref{params_leot} top panel. The mean mass of these dark matter halos (within 300 pc) is $3.7 \times 10^6$ \ms with a standard deviation of $0.7 \times 10^6$ \ms, though the values of $\rho_0$ and $r_s$ vary almost by a factor of three. This, in turn, indicates that there is a strong degeneracy between $\rho_0$ and $r_s$ such that the halo mass is almost constant. Consecutively it implies that the best fit dark matter halo parameters are not unique. In fact, there is a minimal variation in the $\chi^2$ within the hatched region as compared to the outer regions of the parameter space. This leads to the conclusion that within our assumed parameter space, the dark matter halo mass mostly decides the $\Sigma_{HI}$ under hydrostatic equilibrium. As it can be seen from Fig.~\ref{params_leot} top panel, the parameter space of $r_s$ is not fully explored. We restrict our simulation within the shown parameter space, noting that the simulations are computationally expensive and we only wish to provide a rough estimation of the dark matter halo parameters.

\section{Summary and Conclusion}

Using virial theorem, it has long been argued that a typical \HI~cloud with properties of CHVCs/UCHVCc perhaps cannot be stable under self-gravity if they have to be in hydrostatic equilibrium. However, these calculations always assume an isothermal cloud filled with WNM gas. But, there are enough shreds of evidence \citep[see, e.g.,][]{forbq16} that indicate more complex structures in these clouds with an observed velocity dispersion gradient. In these scenarios, proper treatment with a velocity dispersion gradient would be necessary to infer the stability criteria. For example, if we consider Leo T to be a CHVC/UCHVC and has a velocity dispersion $\sim$ 3 \kms~at the centre which increases linearly to a value $\sim 12$ \kms~at its edge, it will have a stable structure {\emph without any dark matter} with $\sim 3$ times the observed \HI~mass in it. This, in turn, will place Leo T at a distance of $\sim$ 720 kpc instead of 420 kpc for the observed \HI~flux which is quite reasonable for a CHVC/UCHVC. Hence, inferring the stability criteria using a simple virial approach with an assumption of isothermal condition might lead to a wrong conclusion. However, as the distance to the UCHVCs/CHVCs are not known, a direct conclusion about the presence of any dark matter in them is not straightforward, though, a limit in the distance can be drawn if one assumes it to be in hydrostatic equilibrium. This can then have significant implications to understand their origin and nature. It should be noted that the absence of any pressure confinement is one of the key assumptions in our calculation. As it has been discussed in details in \citet{faerman13}, the hot-ionized gas around these objects is unlikely to provide enough pressure support to hold these objects as it will require an unnaturally high density of the HIM.

Assuming a prevailing hydrostatic equilibrium, we apply our {\it iterative} method to a gas-rich optically faint Milky Way satellite galaxy Leo T. An assumption of no dark matter results in a very high required \HI~mass or $\Sigma_{HI}$ to maintain its stability. We further use the {\it iterative} method to model the mass distribution in Leo T. We find that a dark matter halo of mass $\sim 2.7 \times 10^6$ \ms~can produce the observed \HI~moment profiles reasonably well. A dark matter halo with core density, $\rho_0 \simeq 0.67$ \mspcc~and a core radius, $r_s \simeq 37$ pc is found to produce a \shi~best matched to the observed one. However, we find a strong degeneracy between these two halo parameters and hence the best fit values are not unique. 

We have also found that all the dark matter halos which can produce the observed $\Sigma_{HI}$ profile within the observed uncertainties have very similar dark matter mass within central 300 pc, even though there is a large variation in $\rho_0$ and $r_s$. This indicates that the \HI~distribution of Leo T is primarily set by the mass of the dark matter halo rather than its detailed structure. However, the structural parameters, e.g., the core density ($\rho_0$) is closely related to the epoch at which the halo formed \citep[see for example][]{strigari08}. Hence, estimating these parameters have cosmological implications. For Leo T, we note that the central core density acquires a flat value of $\sim 0.1$ \mspcc~in halos with large scale-radius. This, in turn, indicates an upper limit of redshift for the formation of the dark matter halo in Leo T. We note that we do not fully explore the parameter space of $\rho_0$ and $r_s$ though, the trend doesn't seem to violet this conclusion.

A similar approach was used by \citet{sternberg02} to estimate the surface density profile of \HI~gas in hydrostatic equilibrium in dark matter halos and was applied to Leo T by \citet{faerman13}. \citet{faerman13} found a dark matter halo mass of $\sim 8 \times 10^6$ \ms~for Leo T within its central 300 pc which is somewhat larger than our result. However, \citet{faerman13} considered the gas to be in isothermal condition, and the gravity due to gas and stars were ignored. Whereas, we self consistently solve the Poisson's equation including gas and stellar mass and allow the velocity dispersion to vary as a function of radius.

\citet{strigari08} performed a Maximum likelihood analysis to estimate the dark matter content of 18 Milky Way satellite galaxies using line-of-sight velocity dispersion data of stars. They determined the dark matter distributions in these galaxies and found halo masses in the range of $3 \times 10^6$ to $3 \times 10^7$ \ms~within the central 300 pc region. Their result indicates a characteristic mass scale of $\sim 10^7$ \ms~in these satellite galaxies. They attributed it to a common mass scale of the dark matter halos which favours the formation of galaxies within them. However, folding into their kinematic information, \citet{collins14} examined the structure of 25 Andromeda dwarf spheroidal galaxies by more than doubling the sample size of previous studies. They found that no universal mass profile can explain the mass distribution in these galaxies which is supported by later numerical studies as well \citep{martinez15}. For Leo T, we also find that no particular mass distribution is uniquely favoured for producing the observed \HI~distribution which support the non-existence of a universal mass profile.

In summary, assuming hydrostatic equilibrium, we set up the Poisson's equation for an \HI~cloud in non-isothermal condition. Using 8$^th$ order Runge-Kutta method, for the first time we solve this equation self-consistently including a stellar component and a dark matter halo. We implement an {\it iterative} method which iteratively estimates the intrinsic velocity dispersion profile suitable to produce the observed moment profiles. By simulating model \HI~clouds and applying the {\it iterative} method, we demonstrate that it can recover the actual moment profiles reasonably well within a maximum error of $\sim 10-15\%$ at the centre. Solving the hydrostatic \HI~structure for Leo T, we also demonstrate that this could be used to infer if an observed \HI~distribution is compilable with hydrostatic equilibrium assumption under self-gravity or not. This, in turn, can be used to identify potential satellite galaxies from a pool of UCHVCs/CHVCs. Further, we model the mass distribution in Leo T by searching through a host of dark matter halos and comparing the modelled $\Sigma_{HI}$ profiles to the observed one. The best fit dark matter halo found to have a $\rho_0 \simeq 0.67$ \mspcc and $r_s \simeq 37$ pc with a mass $\sim 2.7 \times 10^6$ \ms~within central 300 pc. We find a strong degeneracy between the dark matter halo parameters indicating a non-uniqueness of the best fit dark matter halo structure. Interestingly, this also implies that under hydrostatic equilibrium, the mass of a dark matter halo primarily dictates the observed $\Sigma_{HI}$ distribution rather than its detailed structure.

\section{Acknowledgement}
NNP would like to thank Prof. Tom Oosterloo and his team for sharing the \HI~data of Leo T on which the entire work is done. NNP would like to gratefully acknowledge Prof. Jayaram N. Chengalur for his valuable comments and suggestions which has helped to improve the quality of this paper. NNP would like to thank Prof. Marc Verheijen for his suggestions/comments during several rounds of discussions. NNP would like to thank the referee for the constructive comments which helped to improve the quality and the readability of the paper. NNP also would like acknowledge Gunjan Verma for proofreading the manuscript.

\bibliographystyle{mn2e}
\bibliography{bibliography}

\end{document}